\documentclass[conference]{IEEEtran} 
\IEEEoverridecommandlockouts
\usepackage{cite}
\usepackage{amsmath,amssymb,amsfonts}
\usepackage{algorithmic}
\usepackage{graphicx}
\usepackage{textcomp}
\usepackage{xcolor}
\def\BibTeX{{\rm B\kern-.05em{\sc i\kern-.025em b}\kern-.08em
    T\kern-.1667em\lower.7ex\hbox{E}\kern-.125emX}}
\begin{document}

\title{Spatial Temporal Analysis of 40,000,000,000,000 Internet Darkspace Packets
\thanks{This material is based upon work supported by the Assistant Secretary of Defense for Research and Engineering under Air Force Contract No. FA8702-15-D-0001, National Science Foundation CCF-1533644, and United States Air Force Research Laboratory Cooperative Agreement Number FA8750-19-2-1000. Any opinions, findings, conclusions or recommendations expressed in this material are those of the author(s) and do not necessarily reflect the views of the Assistant Secretary of Defense for Research and Engineering, the National Science Foundation, or the United States Air Force. The U.S. Government is authorized to reproduce and distribute reprints for Government purposes notwithstanding any copyright notation herein.}
}

\author{\IEEEauthorblockN{Jeremy Kepner$^1$, Michael Jones$^1$, Daniel Andersen$^2$, Ayd{\i}n Bulu{\c{c}}$^3$, Chansup Byun$^1$,   K Claffy$^2$, Timothy Davis$^4$,  \\ William Arcand$^1$, Jonathan Bernays$^1$, David Bestor$^1$, William Bergeron$^1$, Vijay Gadepally$^1$,  \\ Micheal Houle$^1$, Matthew Hubbell$^1$,  Anna Klein$^1$, Chad Meiners$^1$, Lauren Milechin$^1$, Julie Mullen$^1$, \\ Sandeep Pisharody$^1$, Andrew Prout$^1$,  Albert Reuther$^1$, Antonio Rosa$^1$, Siddharth Samsi$^1$, \\ Doug Stetson$^1$, Adam Tse$^1$, Charles Yee$^1$, Peter Michaleas$^1$
\\
\IEEEauthorblockA{$^1$MIT,  $^2$CAIDA, $^3$LBNL, $^4$Texas A\&M
}}}
\maketitle

\begin{abstract}
The Internet has never been more important to our society, and understanding the behavior of the Internet is essential.  The Center for Applied Internet Data Analysis (CAIDA) Telescope observes a continuous stream of packets from an unsolicited darkspace representing 1/256 of the Internet.  During 2019 and 2020 over 40,000,000,000,000 unique packets were collected representing the largest ever assembled public corpus of Internet traffic.  Using the combined resources of the Supercomputing Centers at UC San Diego, Lawrence Berkeley National Laboratory, and MIT, the spatial temporal structure of anonymized source-destination pairs from the CAIDA Telescope data has been analyzed with GraphBLAS hierarchical hypersparse matrices.  These analyses provide unique insight on this unsolicited Internet darkspace traffic with the discovery of many previously unseen scaling relations.  The data show a significant sustained increase in unsolicited traffic corresponding to the start of the COVID19 pandemic, but relatively little change in the underlying scaling relations associated with unique sources, source fan-outs, unique links, destination fan-ins, and unique destinations.  This work provides a demonstration of the practical feasibility and benefit of the safe collection and analysis of significant quantities of anonymized Internet traffic.  
\end{abstract}


\section{Introduction}
\begin{figure}
\center{\includegraphics[width=1.0\columnwidth]{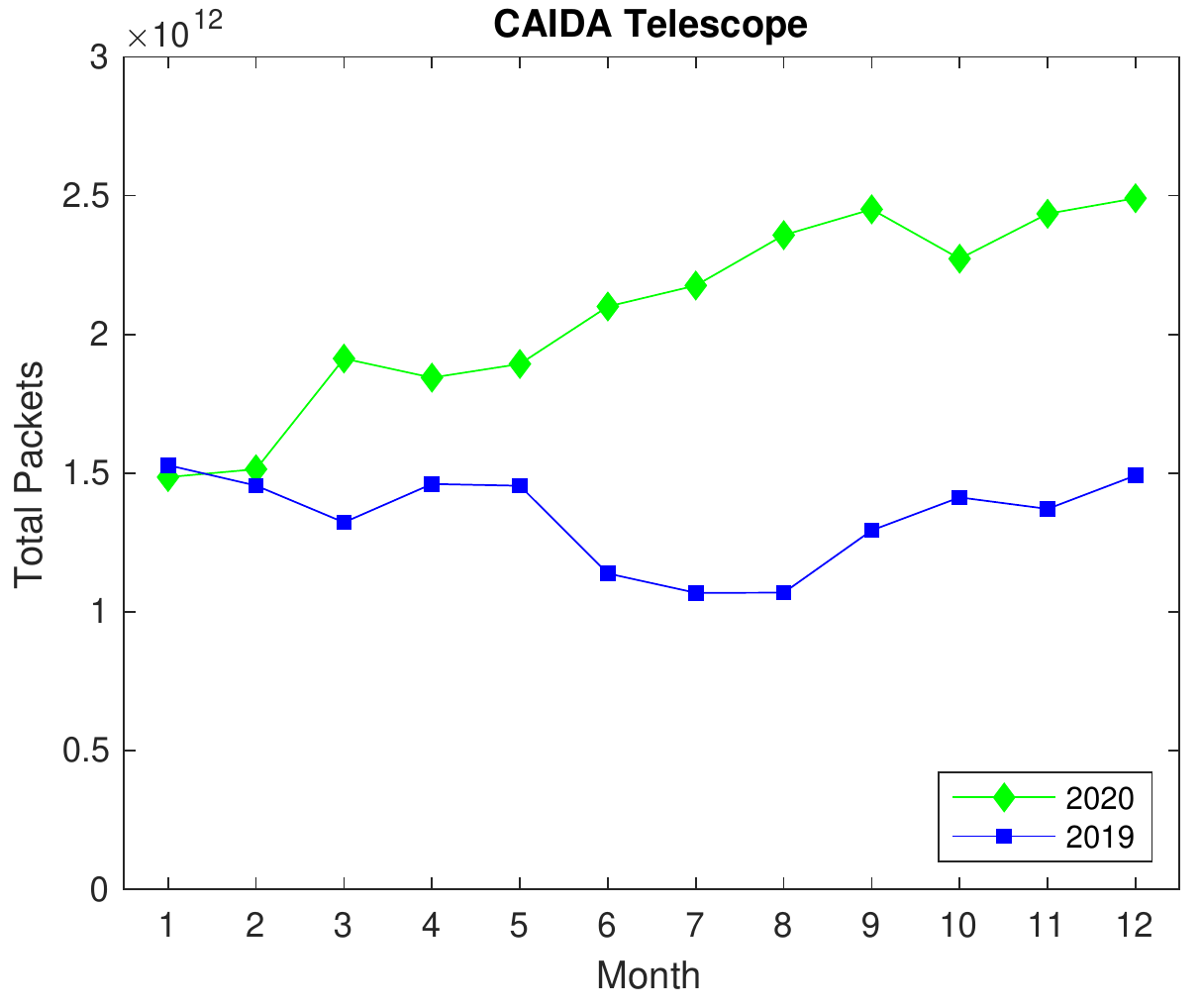}}
      	\caption{{\bf Packets per month.} Number of packets collected per month by the CAIDA telescope. The number of packets increased significantly during the COVID19 pandemic.}
      	\label{fig:TotalPackets}
\end{figure}

  For over five billion people the Internet has become as essential as land, sea, air, and space for enabling activities as diverse as commerce, education, health, and entertainment \cite{Cisco2018-2023}.  Understanding the Internet is likewise as important as studying these other domains \cite{kepner2021zero}.  Developing scientific insights on how the Internet behaves requires observation and data \cite{claffy2000measuring, li2013survey, rabinovich2016measuring, ClaffyClark2020}.  The largest public Internet observatory is the Center for Applied Internet Data Analysis (CAIDA) Telescope that operates a variety of sensors including a continuous stream of packets from an unsolicited darkspace representing 1/256 of the Internet.   This network telescope monitors an Internet darkspace (also referred to as a black hole, Internet sink, or darknet) that is a globally routed /8 network that carries almost no legitimate traffic because there are few allocated addresses in this Internet prefix. After discarding the small amount of legitimate traffic from the incoming packets, the remaining data represent a continuous view of anomalous unsolicited traffic, or Internet background radiation. Almost every computer on the Internet will receive some form of this background traffic.  This unsolicited traffic results from a wide range of events, such as backscatter from randomly spoofed source denial-of-service attacks, the automated spread of Internet worms and viruses, scanning of address space by attackers or malware looking for vulnerable targets, and various misconfigurations (e.g. mistyping an IP address). In recent years, traffic destined to darkspace has evolved to include longer-duration, low-intensity events intended to establish and maintain botnets.  CAIDA personnel maintain and expand the telescope instrumentation, collects, curates, archives, and analyzes the data, and enables data access for vetted  researchers.

During 2019 and 2020 over 40,000,000,000,000 unique packets were collected by the CAIDA Telescope.  This data set represents the largest ever assembled public corpus of Internet traffic, and is perhaps the largest public collection of streaming network events of any type.  Figure~\ref{fig:TotalPackets} shows the number of packet in each month and indicates a significant increase aligning with the COVID19 pandemic.  Analysis of such a large network data set is computationally challenging \cite{lumsdaine2007challenges, kolda2009tensor, hilbert2011world}.  Using the combined resources of the Supercomputing Centers at UC San Diego, Lawrence Berkeley National Laboratory, and MIT, the spatial temporal structure of anonymized source-destination pairs from the CAIDA Telescope data has been analyzed leveraging prior work on massively parallel GraphBLAS hierarchical hypersparse matrices \cite{Kepner2009, kepner2011graph, kepner2018mathematics, reuther2018interactive, gadepally2018hyperscaling, kepner19streaming, kepner202075}.  Applying this type of analysis to other collections of billions of network packets has revealed power-law distributions \cite{kepner19hypersparse}, novel scaling relations \cite{kepner2020multi}, and inspired new models of network traffic \cite{devlin2021hybrid}.  The goal of this paper is to understand the scaling relations revealed by the  CAIDA telescope data set.  These scaling relations can provide fundamental insights into Internet background traffic. This work can also provide a demonstration of the practical feasibility and benefit of the safe collection and analysis of significant of quantities anonymized Internet traffic.

The outline of the rest of the paper is as follows.  First, the network quantities and their distributions are defined in terms of traffic matrices.   Second, multi-temporal analysis of hypersparse hierarchical traffic matrices is described.  Third, the method for determining scaling relations as a function of the packet window $N_V$ is presented along with the resulting scaling relations observed in the gateway traffic data.  Finally, our conclusions and directions for further work are presented.

\section{Network Quantities and Distributions from Traffic Matrices}

\begin{figure}
\center{\includegraphics[width=0.65\columnwidth]{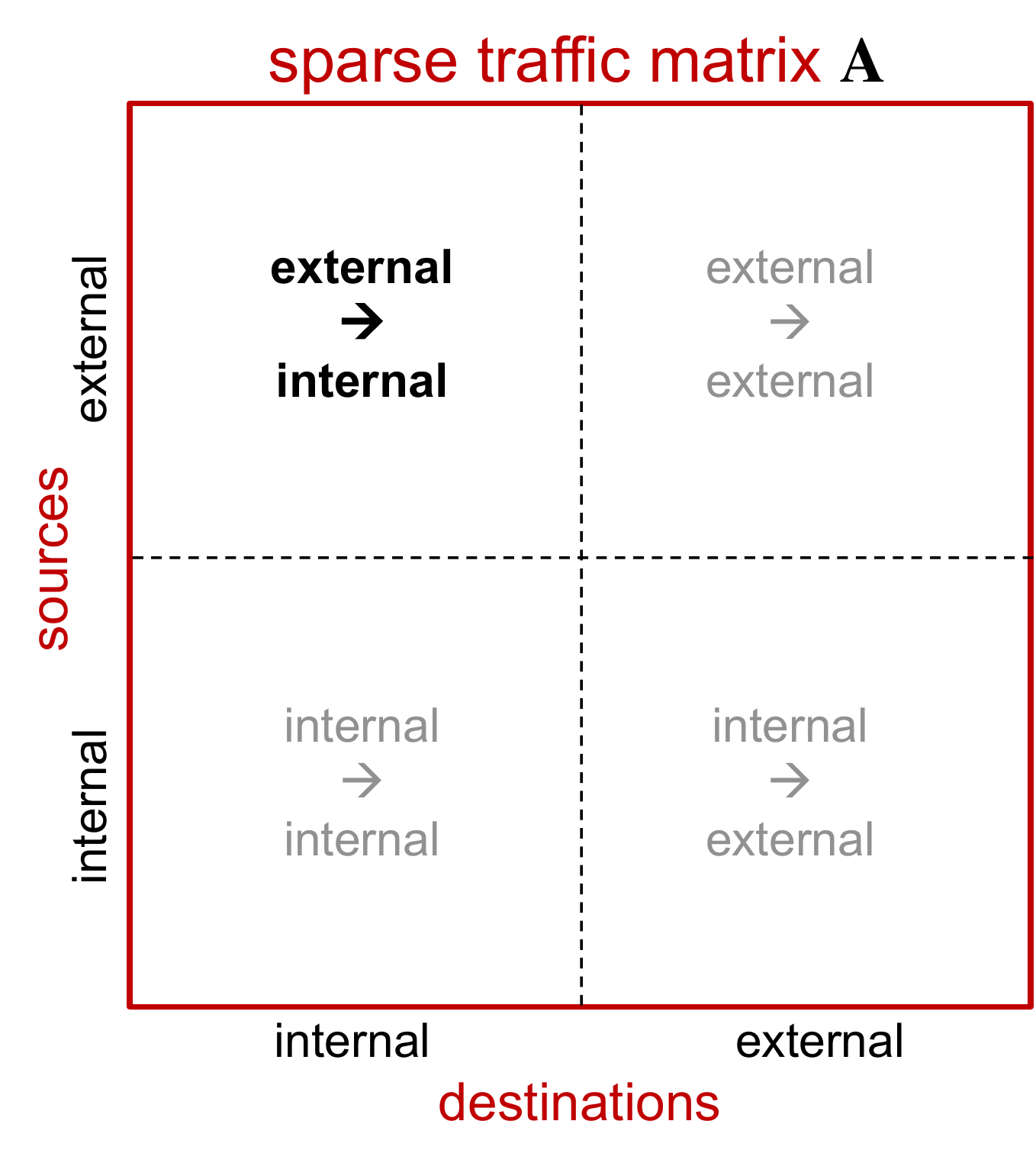}}
      	\caption{{\bf Network traffic matrix.} The traffic matrix can be divided into quadrants separating internal and external traffic.  The CAIDA Telescope monitors a darkspace, so only the upper left (external $\rightarrow$ internal) quadrant will have data.}
      	\label{fig:GatewayTrafficMatrix}
\end{figure}

The CAIDA Telescope monitors the traffic into and out of a set of network addresses providing a natural observation point of network traffic.  These data can be viewed as a traffic matrix where each row is a source and each column is a destination.  The CAIDA Telescope traffic matrix can be partitioned into four quadrants (see Figure~\ref{fig:GatewayTrafficMatrix}).  These quadrants represent different flows between nodes internal and external to the set of monitored addresses.  Because the CAIDA Telescope network addresses are a darkspace, only the upper left (external $\rightarrow$ internal) quadrant will have data.   Internet data  must be handled with care, and CAIDA has pioneered standard trusted data sharing best practices that include \cite{kepner2021zero}
\begin{itemize}
\item Data is made available in curated repositories
\item Using standard anonymization methods where needed: hashing, sampling, and/or simulation
\item Registration with a repository and demonstration of legitimate research need
\item Recipients legally agree to neither repost a corpus nor deanonymize data
\item Recipients can publish analysis and data examples necessary to review research
\item Recipients agree to cite the repository and provide publications back to the repository
\item Repositories can curate enriched products developed by researchers
\end{itemize}

\begin{figure}
\center{\includegraphics[width=1.0\columnwidth]{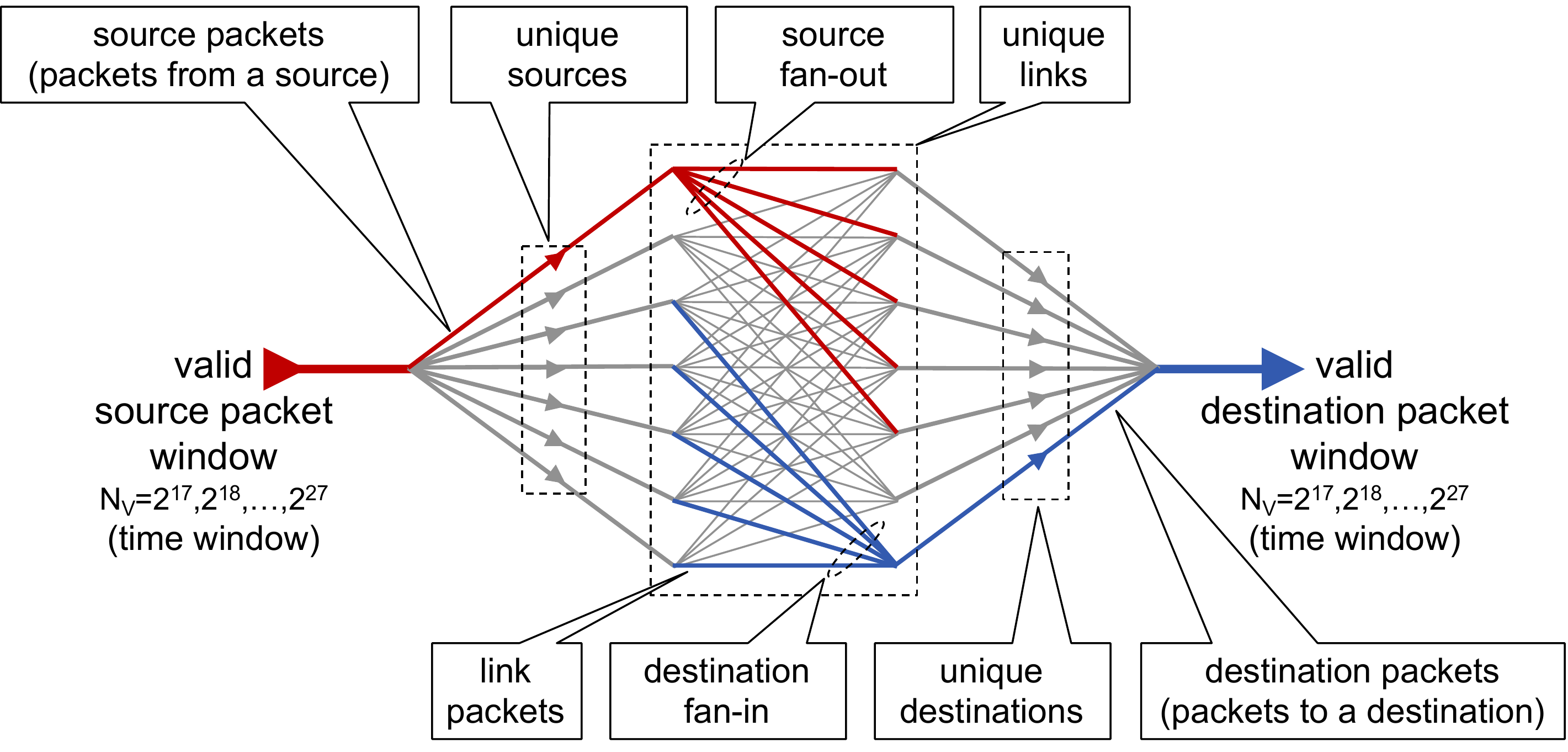}}
      	\caption{{\bf Streaming network traffic quantities.} Internet traffic streams of $N_V$ valid packets are divided into a variety of quantities for analysis: source packets, source fan-out, unique source-destination pair packets (or links), destination fan-in, and destination packets.}
      	\label{fig:NetworkDistribution}
\end{figure}

  Streams of interactions between entities are found in many domains.  For Internet traffic these interactions are referred to as packets \cite{huang2018software}.  Figure~\ref{fig:NetworkDistribution} illustrates essential quantities found in all streaming dynamic networks. These quantities are all computable from anonymized traffic matrices created from the source and destinations found in packet headers.  These sources and destinations are referred as Internet Protocol (IP) addresses.  

Processing such a large volume of data requires additional computational innovations. The advent of GraphBLAS hypersparse hierarchical traffic matrices has enabled the processing of hundreds of billions of packets in minutes \cite{kepner16mathematical, buluc17design, davis18algorithm, kepner202075}.
The CAIDA Telescope archives its trillions of collected packets at the supercomputing center at Lawrence Berkeley National Laboratory (LBNL) where the packets are aggregated into CryptoPAN \cite{fan2004prefix} anonymized GraphBLAS traffic matrices of $N_V = 2^{17}$ contiguous packets.  This process compresses the data down to approximately 3 bytes/packet.   The resulting matrices are stored and sent to the MIT SuperCloud where the network quantities shown in Figure~\ref{fig:NetworkDistribution} are computed.  Using 384 64-core compute nodes (24,576 cores total) on the MIT SuperCloud all 40,000,000,000,000 packets were processed in four days.  

The code was implemented using Matlab/Octave with the pMatlab parallel library \cite{Kepner2009}.  A typical run could be launched in a few seconds using the MIT SuperCloud triples-mode hierarchical launching system \cite{reuther2018interactive}.  Typical launch parameters were [384 16 4], corresponding to 384 nodes, 16 Matlab/Octave processes per node, and 4 OpenMP threads per process.  On each node, the 16 processes were pinned to 4 adjacent cores to minimize interprocess contention and maximize cache locality for the GraphBLAS OpenMP threads \cite{byun2019optimizing}.  Three levels of parallelism were used within the program.  At the top level each month in a year was processed independently using 384/12 = 32 compute nodes.  Within each month, the packet windows were were split among the 32$\times$16 = 512 Matlab/Octave processes with some overlap to allow for contiguous processing of the streaming data.  Within each Matlab/Octave process, the underlying GraphBLAS OpenMP parallelism was used on 4 cores.  At the end of the processing the results were aggregated using asynchronous file-based messaging \cite{byun2019large}.

The contiguous nature of these data allows the exploration of a wide range of packet windows from $N_V = 2^{17}$ (sub-second) to $N_V = 2^{27}$ (minutes), providing a unique view into how network quantities depend upon time.  These observations provide new insights into normal network background traffic that could be used for anomaly detection, AI feature engineering, polystore index learning, and testing theoretical models of streaming networks \cite{elmore2015demonstration, kraska18case, do20classifying}.

The network quantities depicted in Figure~\ref{fig:NetworkDistribution} are computable from anonymized origin-destination  matrices that are widely used to represent network traffic \cite{soule2004identify, zhang2005estimating, mucha2010community, tune2013internet}.  It is common to filter the packets down to a valid set for  any particular analysis.   Such filters may limit particular sources, destinations, protocols, and time windows. To reduce statistical fluctuations, the streaming data should be partitioned so that for any chosen time window all data sets have the same number of valid packets \cite{kepner19streaming}.  At a given time $t$, $N_V$ consecutive valid packets are aggregated from the traffic into a sparse matrix ${\bf A}_t$, where ${\bf A}_t(i,j)$ is the number of valid packets between the source $i$ and destination $j$. The sum of all the entries in ${\bf A}_t$ is equal to $N_V$
$$
    \sum_{i,j} {\bf A}_t(i,j) = N_V
$$
All the network quantities depicted in Figure~\ref{fig:NetworkDistribution} can be readily computed from ${\bf A}_t$ using the formulas listed in Table~\ref{tab:Aggregates}.  Because matrix operations are generally invariant to permutation (reordering of the rows and columns), these quantities can readily be computed from anonymized data.

\begin{table}
\caption{Network Quantities from Traffic Matrices}
\vspace{-0.25cm}
Formulas for computing network quantities from  traffic matrix ${\bf A}_t$ at time $t$ in both summation and matrix notation. ${\bf 1}$ is a column vector of all 1's, $^{\sf T}$  is the transpose operation, and $|~|_0$ is the zero-norm that sets each nonzero value of its argument to 1\cite{karvanen2003measuring}.  These formulas are unaffected by matrix permutations and will work on anonymized data.
\begin{center}
\begin{tabular}{p{1.45in}p{0.9in}p{0.6in}}
\hline
{\bf Aggregate} & {\bf ~~~~Summation} & {\bf ~Matrix} \\
{\bf Property} & {\bf ~~~~~~Notation} & {\bf Notation} \\
\hline
Valid packets $N_V$ & $~~\sum_i ~ \sum_j ~ {\bf A}_t(i,j)$ & $~{\bf 1}^{\sf T} {\bf A}_t {\bf 1}$ \\
Unique links & $~~\sum_i ~ \sum_j |{\bf A}_t(i,j)|_0$  & ${\bf 1}^{\sf T}|{\bf A}_t|_0 {\bf 1}$ \\
Link packets from $i$ to $j$ & $~~~~~~~~~~~~~~{\bf A}_t(i,j)$ & ~~~$~{\bf A}_t$ \\
Max link packets ($d_{\rm max}$) & $~~~~~\max_{ij}{\bf A}_t(i,j)$ & $\max({\bf A}_t)$ \\
\hline
Unique sources & $~\sum_i |\sum_j ~ {\bf A}_t(i,j)|_0$  & ${\bf 1}^{\sf T}|{\bf A}_t {\bf 1}|_0$ \\
Packets from source $i$ & $~~~~~~~\sum_j ~ {\bf A}_t(i,j)$ & ~~$~~{\bf A}_t  {\bf 1}$ \\
Max source packets ($d_{\rm max}$)  & $ \max_i \sum_j ~ {\bf A}_t(i,j)$ & $\max({\bf A}_t {\bf 1})$ \\
Source fan-out from $i$ & $~~~~~~~~~~\sum_j |{\bf A}_t(i,j)|_0$  & ~~~$|{\bf A}_t|_0 {\bf 1}$ \\
Max source fan-out ($d_{\rm max}$) & $ \max_i \sum_j |{\bf A}_t(i,j)|_0$  & $\max(|{\bf A}_t|_0 {\bf 1})$ \\
\hline
Unique destinations & $~\sum_j |\sum_i ~ {\bf A}_t(i,j)|_0$ & $|{\bf 1}^{\sf T} {\bf A}_t|_0 {\bf 1}$ \\
Destination packets to $j$ & $~~~~~~~\sum_i ~ {\bf A}_t(i,j)$ & ${\bf 1}^{\sf T}|{\bf A}_t|_0$ \\
Max destination packets ($d_{\rm max}$) & $ \max_j \sum_i ~ {\bf A}_t(i,j)$ & $\max({\bf 1}^{\sf T}|{\bf A}_t|_0)$ \\
Destination fan-in to $j$ & $~~~~~~~~~~\sum_i |{\bf A}_t(i,j)|_0$ & ${\bf 1}^{\sf T}~{\bf A}_t$ \\
Max destination fan-in ($d_{\rm max}$) & $ \max_j \sum_i |{\bf A}_t(i,j)|_0$ & $\max({\bf 1}^{\sf T}~{\bf A}_t)$ \\
\hline
\end{tabular}
\end{center}
\label{tab:Aggregates}
\end{table}%

\begin{figure}
\includegraphics[width=0.5\columnwidth]{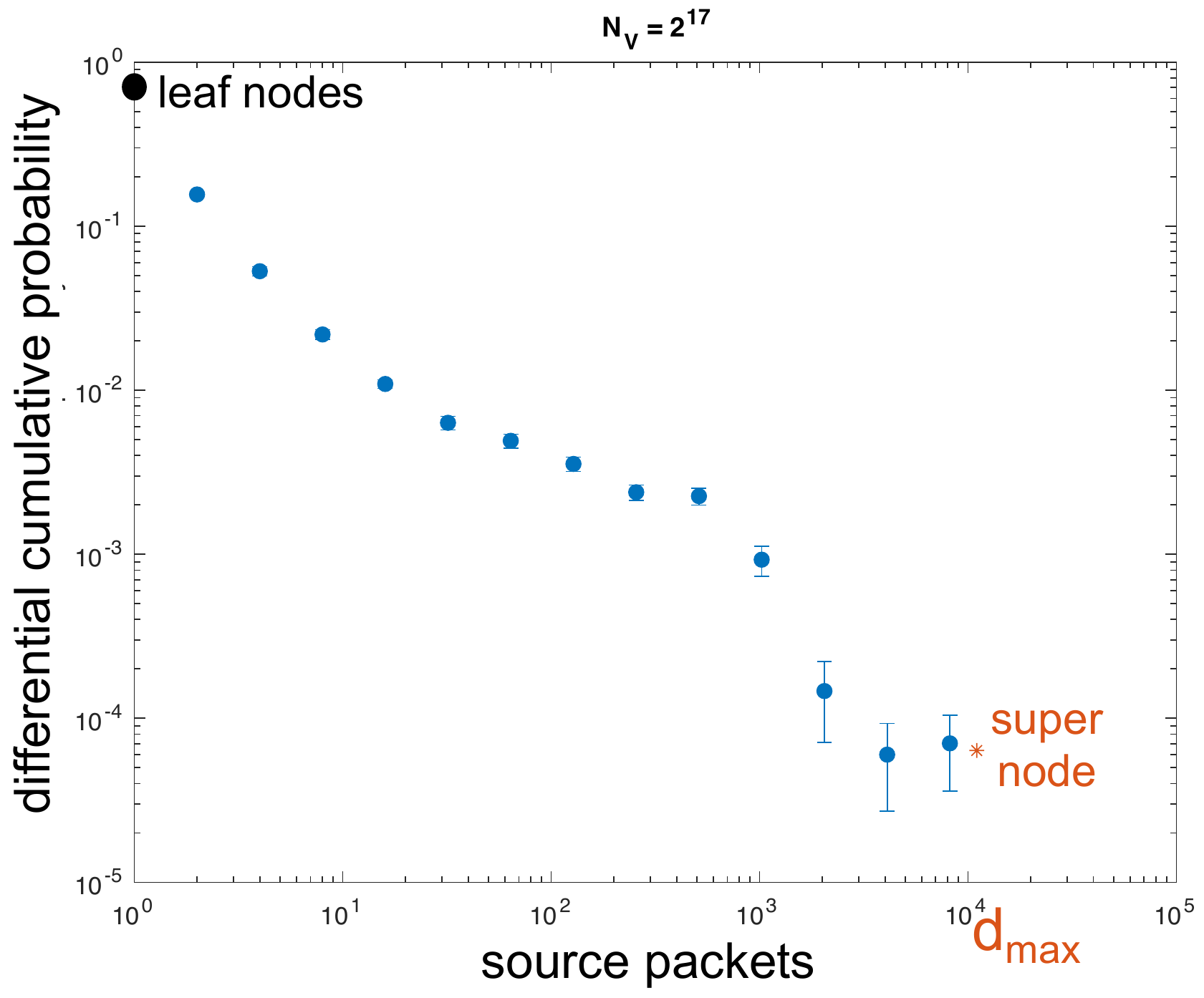}\includegraphics[width=0.5\columnwidth]{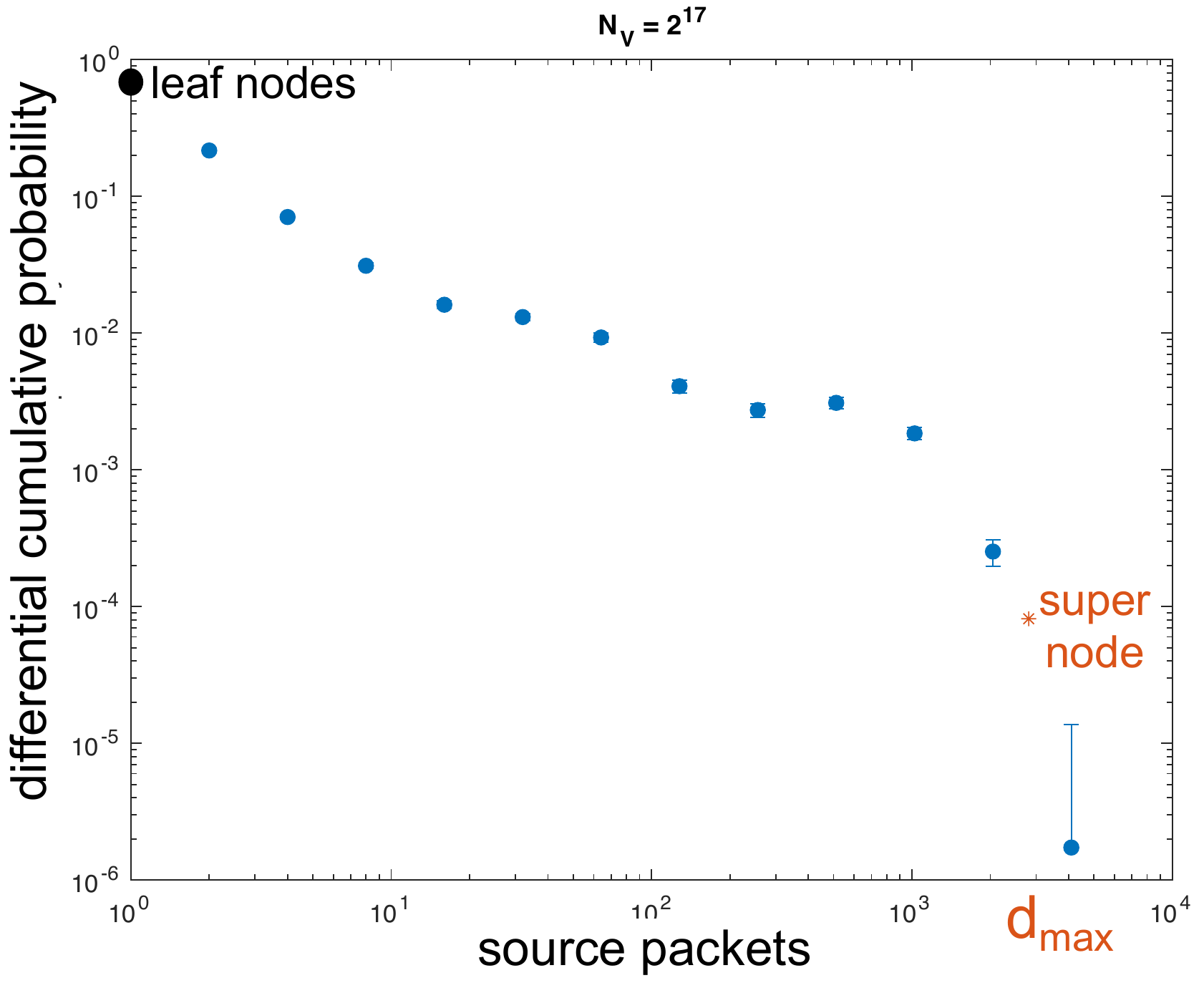}
\includegraphics[width=0.5\columnwidth]{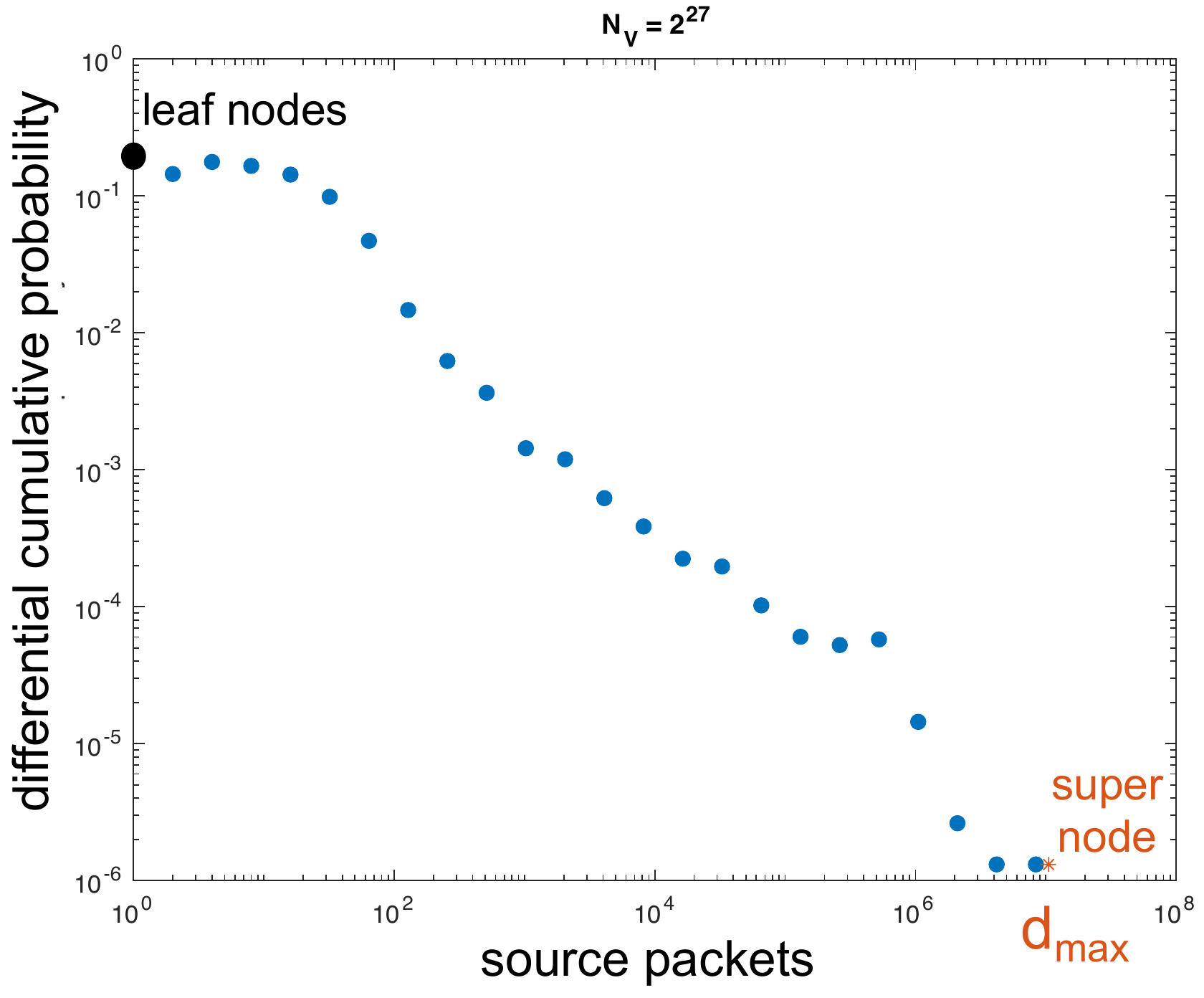}\includegraphics[width=0.5\columnwidth]{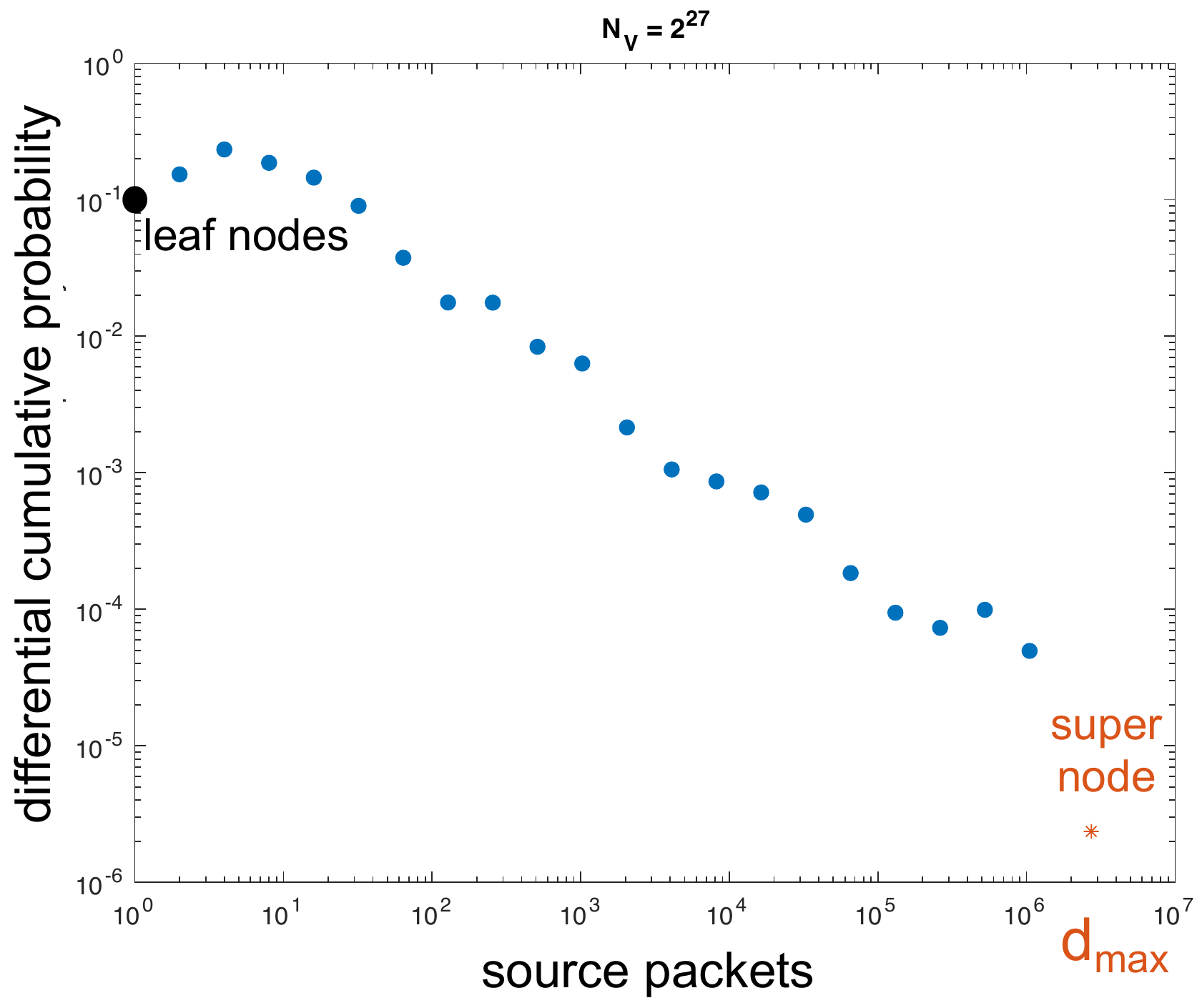}
      	\caption{{\bf Source packet degree distribution} (left 2019-06, right 2020-06).  Example differential cumulative probability (normalized histogram) for two packet windows ($N_V=2^{17}$ and $N_V=2^{27}$) of the number (degree) of source packets from each source using logarithmic bins $d_i = 2^i$.  Sources sending out a single packet are denoted ``leaf nodes''.  The source with the largest number of packets $d_{\rm max}$ is referred to as the ``supernode''.}
      	\label{fig:DegreeDistribution}
\end{figure}

Each network quantity will produce a distribution of values whose magnitude is often called the degree $d$. The corresponding histogram of the network quantity is denoted by $n_t(d)$.  The largest observed value in the distribution is denoted  $d_{\rm max}$.  The normalization factor of the distribution is given by
$$
    \sum_d n_t(d)
$$
with corresponding probability
$$
    p_t(d) = n_t(d)/\sum_d n_t(d)
$$
and cumulative probability
$$
    P_t(d) = \sum_{i=1,d} p_t(d)
$$
Because of the relatively large values of $d$ observed, the measured probability at large $d$ often exhibits large fluctuations. However, the cumulative probability lacks sufficient detail to see variations around specific values of $d$, so it is typical to pool the \emph{differential cumulative probability} with logarithmic bins in $d$
$$
    D_t(d_i) = P_t(d_i) - P_t(d_{i-1})
$$
where $d_i = 2^i$ \cite{clauset2009power}.  All computed probability distributions use the same binary logarithmic binning to allow for consistent statistical comparison across data sets \cite{clauset2009power, barabasi2016network}.  The corresponding mean and standard deviation of $D_t(d_i)$ over many different consecutive values of $t$ for a given data set are denoted $D(d_i)$ and $\sigma(d_i)$. Figure~\ref{fig:DegreeDistribution} provides an example distribution of external $\rightarrow$ internal source packets with packet windows of $N_V = 2^{17}$ and $N_V = 2^{27}$ at two different times.  The resulting distribution exhibits the power-law frequently observed in network  data  \cite{leland1994self, faloutsos1999power, albert1999internet, barabasi1999emergence, adamic2000power, barabasi2009scale, mahanti2013tale}.

\section{Multi-Temporal Analysis}

\begin{figure}
\center{\includegraphics[width=1.0\columnwidth]{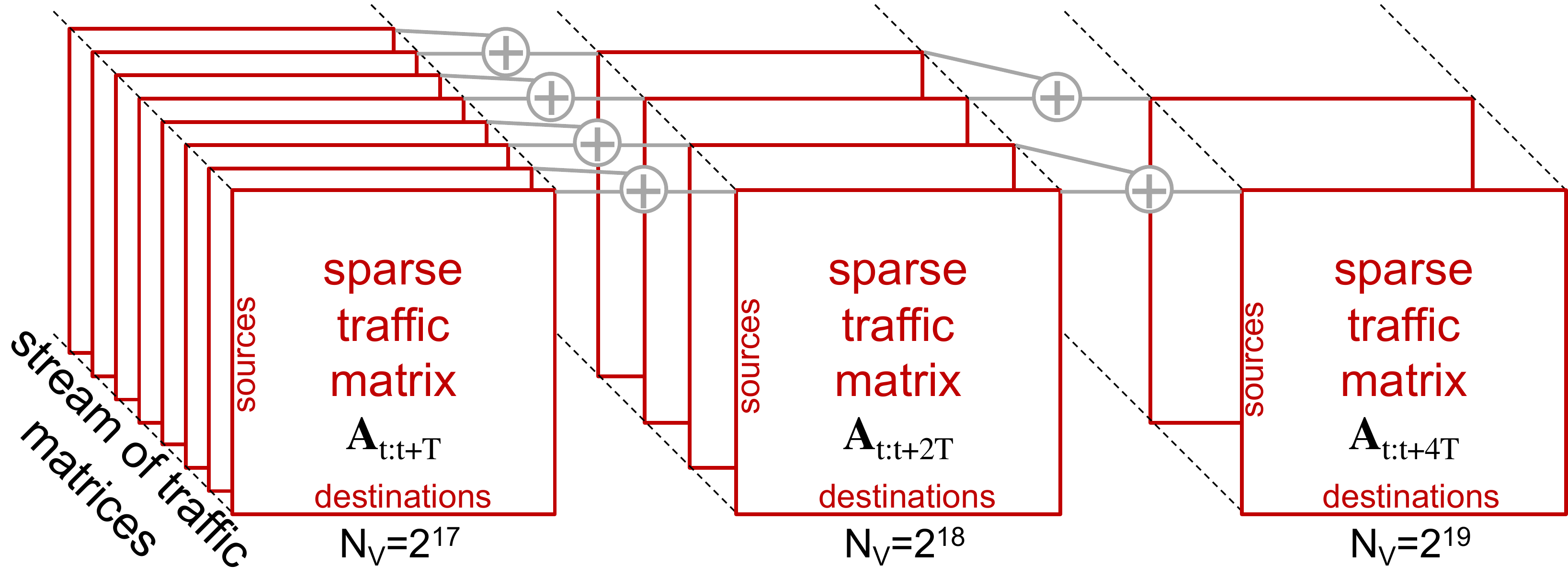}}
      	\caption{{\bf Multi-temporal streaming traffic matrices.} Efficient computation of network quantities on multiple time scales can be achieved by hierarchically aggregating data in different time windows.}
      	\label{fig:MultiTemporalMatrix}
\end{figure}

Network traffic is dynamic and exhibits varying behavior on a wide range of time scales.  A given packet window size $N_V$ will be sensitive to phenomena on its corresponding timescale.  Determining how network quantities scale with $N_V$ provides insight into the temporal behavior of network traffic.   Efficient computation of network quantities on multiple time scales can be achieved by hierarchically aggregating data in different time windows \cite{kepner19streaming}.  Figure~\ref{fig:MultiTemporalMatrix} illustrates a binary aggregation of  different streaming traffic matrices.   Computing each quantity at each hierarchy level eliminates redundant computations that would be performed if each packet window was computed separately.  Hierarchy also ensures that most computations are performed on smaller matrices residing in faster memory.  Correlations among the matrices mean  that adding two matrices each with $N_V$ entries results in a matrix with fewer than $2N_V$ entries, reducing the relative number of operations as the matrices grow.

One of the important capabilities of the SuiteSparse GraphBLAS library is support for hypersparse matrices where the number of nonzero entries is significantly less than either dimensions of the matrix.  If the packet source and destination identifiers are drawn from a large numeric range, such as those used in the Internet protocol, then a hypersparse representation of ${\bf A}_t$ eliminates the need to keep track of additional indices and can significantly accelerate the computations \cite{kepner202075}.

\begin{figure*}
\includegraphics[width=\columnwidth]{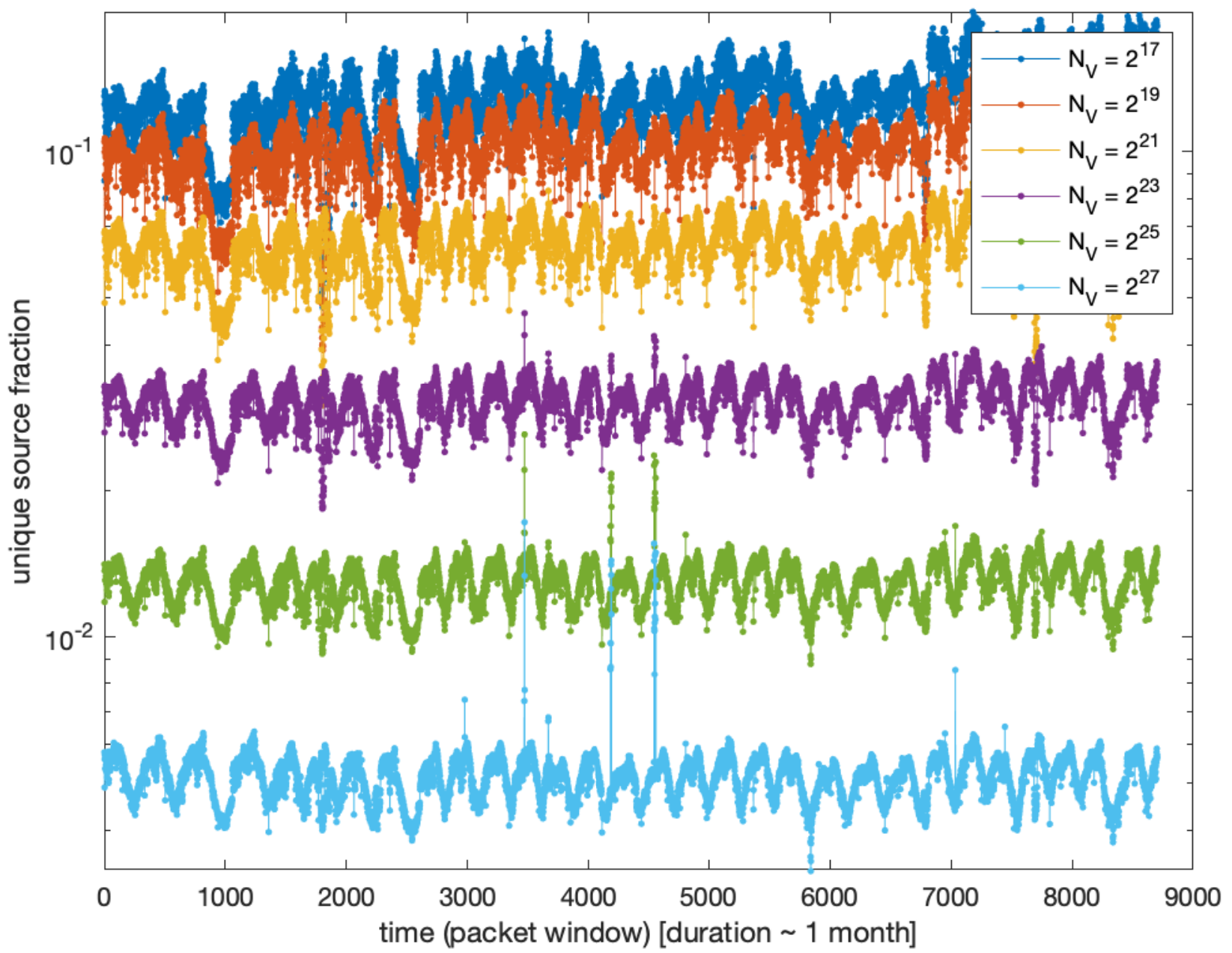}\includegraphics[width=\columnwidth]{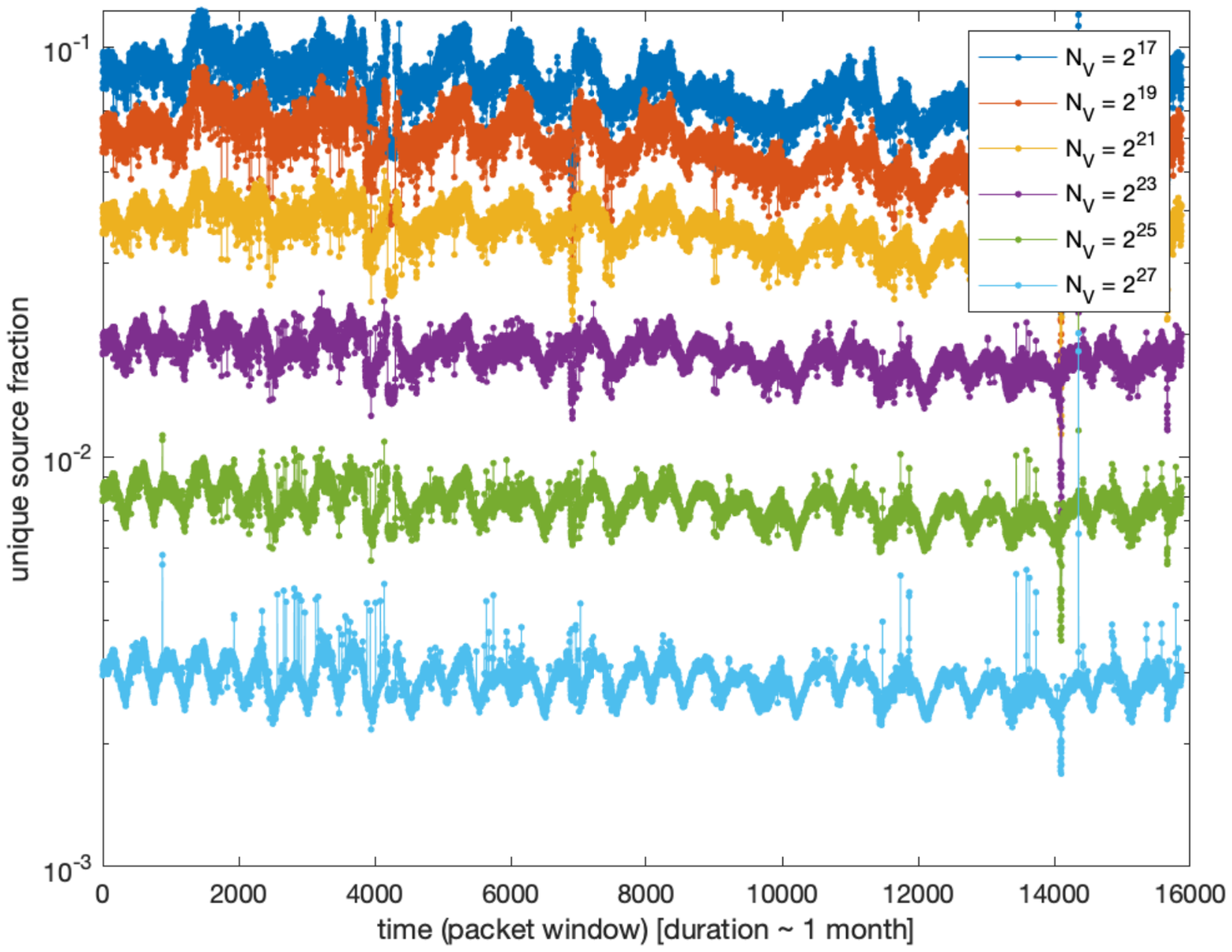}
\includegraphics[width=\columnwidth]{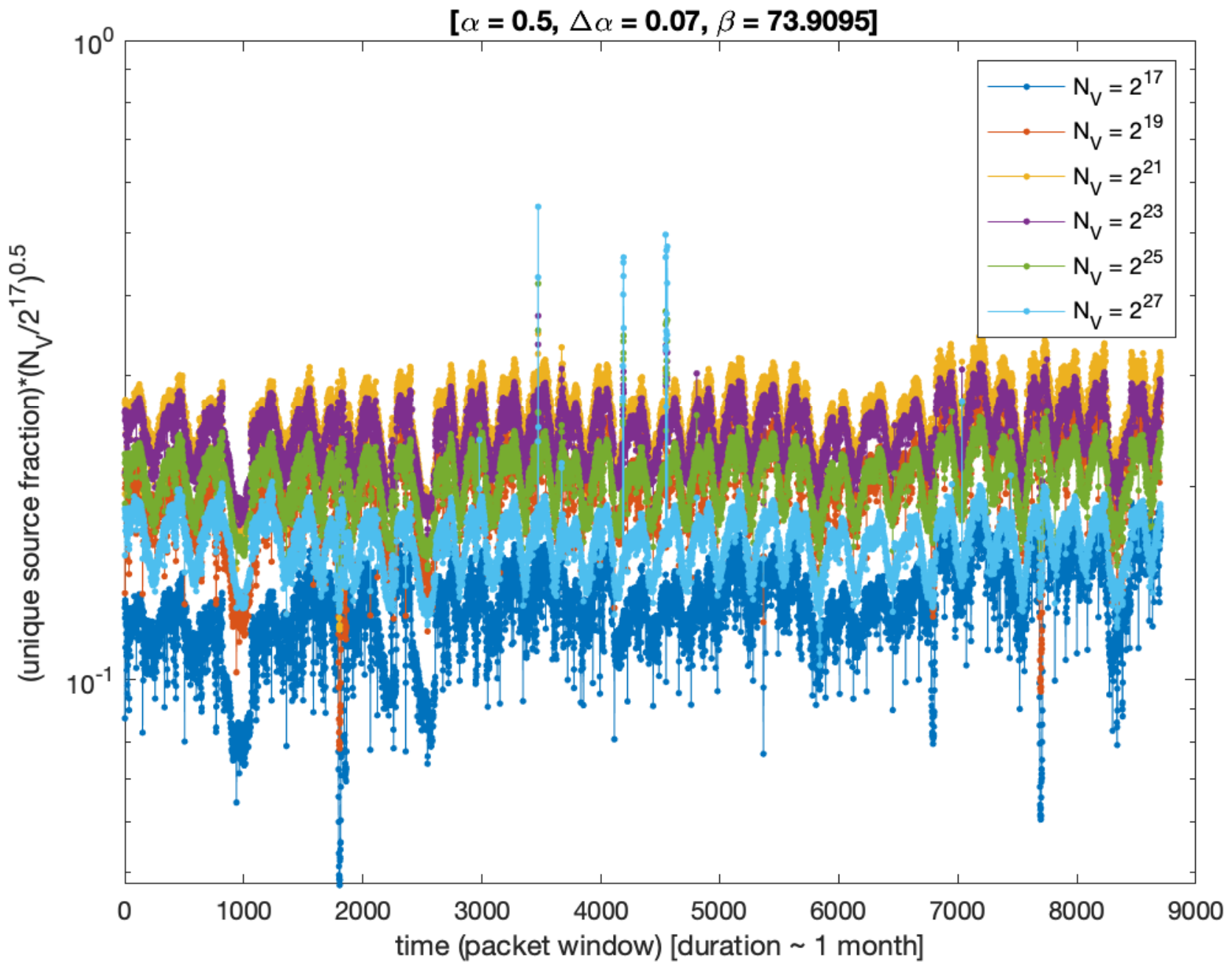}\includegraphics[width=\columnwidth]{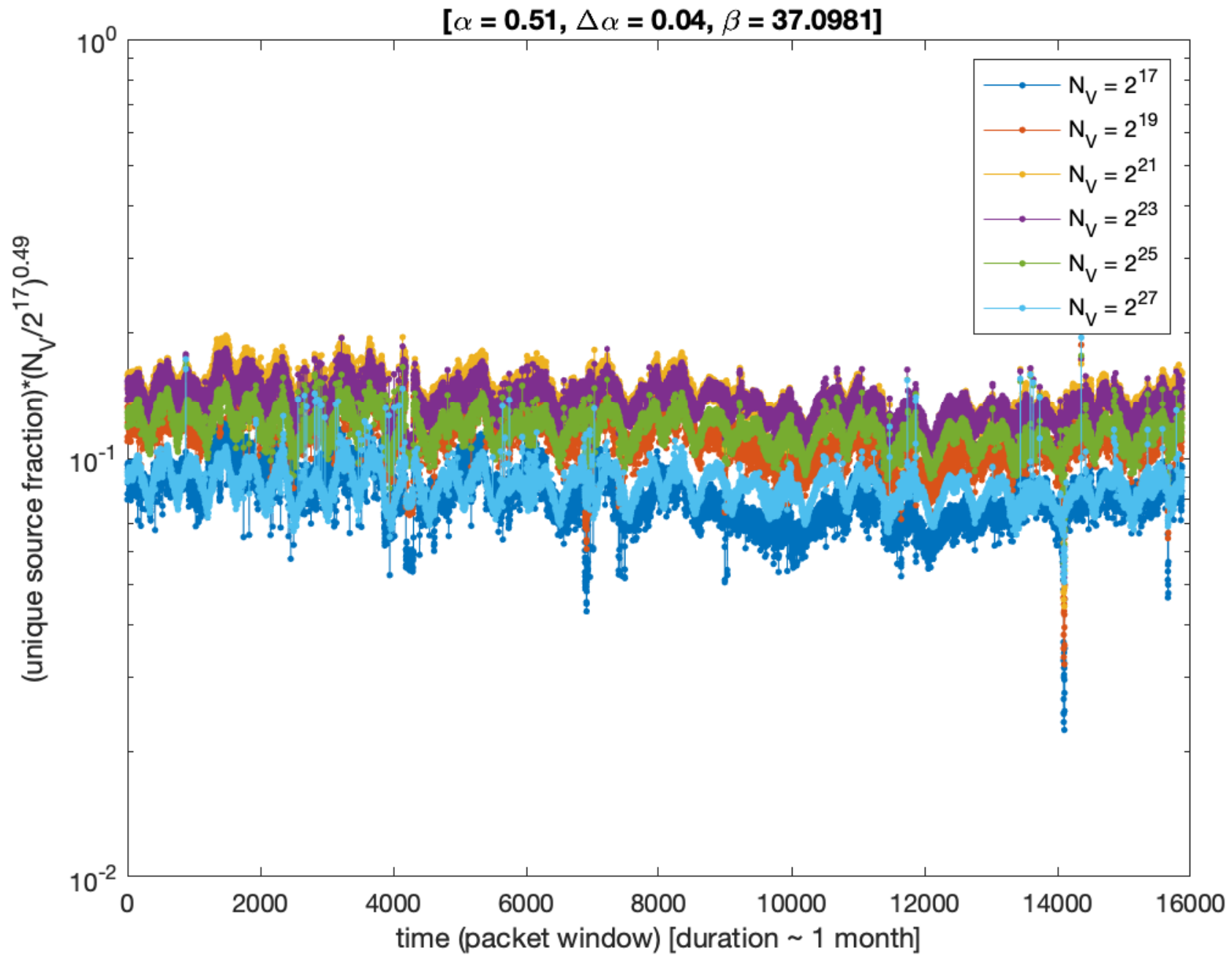}
      	\caption{{\bf (top) Unique source fraction} (left 2019-06, right 2020-06).  Average total number of unique sources in a packet window of width $N_V$ measured at each time over a month.  {\bf (bottom) Normalized unique source fraction} (left 2019-06, right 2020-06).  Scaling (top) data by $(N_V/2^{17})^{0.5}$ and $(N_V/2^{17})^{0.49}$ aligns the different packet windows, indicating that the number of uniques sources is proportional to $N_V^{1-0.5} = N_V^{0.5}$ and $N_V^{1-0.49} = N_V^{0.51}$.  $\Delta \alpha$ is the difference between the best fit $\alpha$ obtained using the different error norms $|~|^2$, $|~|$, and $|~|_0$}
      	\label{fig:UniqueSourceFraction}
\end{figure*}

\begin{table*}[htp]
\caption{Approximate scaling relations.}
\vspace{-0.25cm}
Analysis of network quantities over packet windows $N_V = 2^{17}, \ldots, 2^{27}$ reveals a strong dependence of many of these quantities on $N_V$ as well as remarkable consistency between 2019 and 2020.\center{\includegraphics[width=1.75\columnwidth]{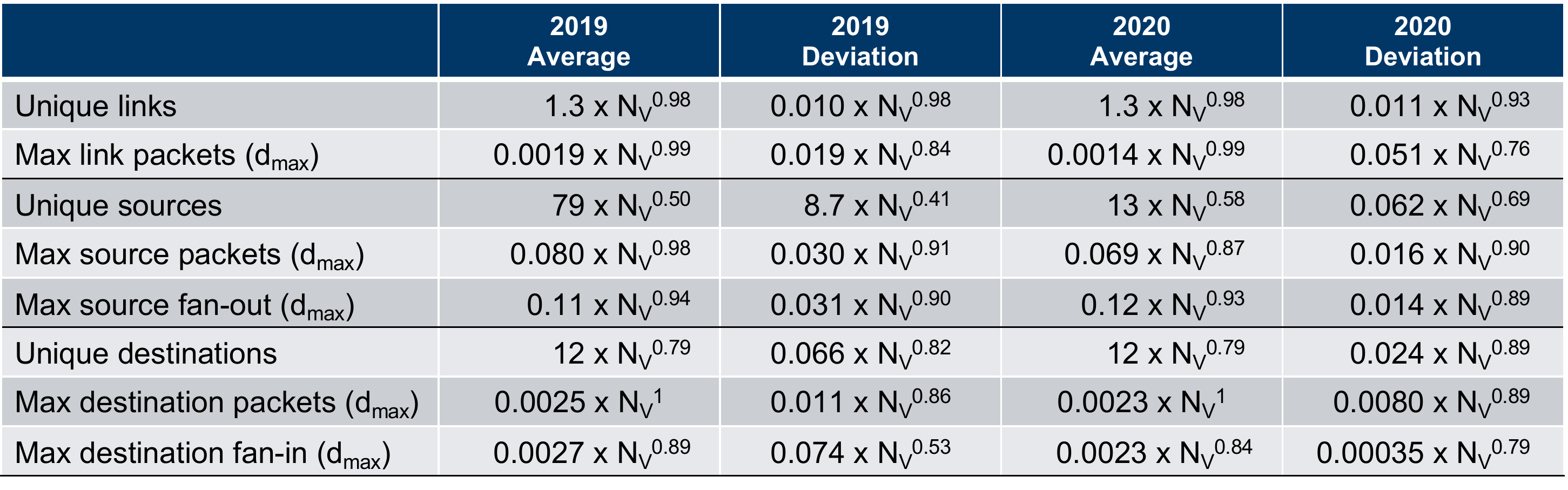}}
\label{tab:ScalingRelations}
\end{table*}%

\section{Results}

\begin{figure}
\center{\includegraphics[width=0.8\columnwidth]{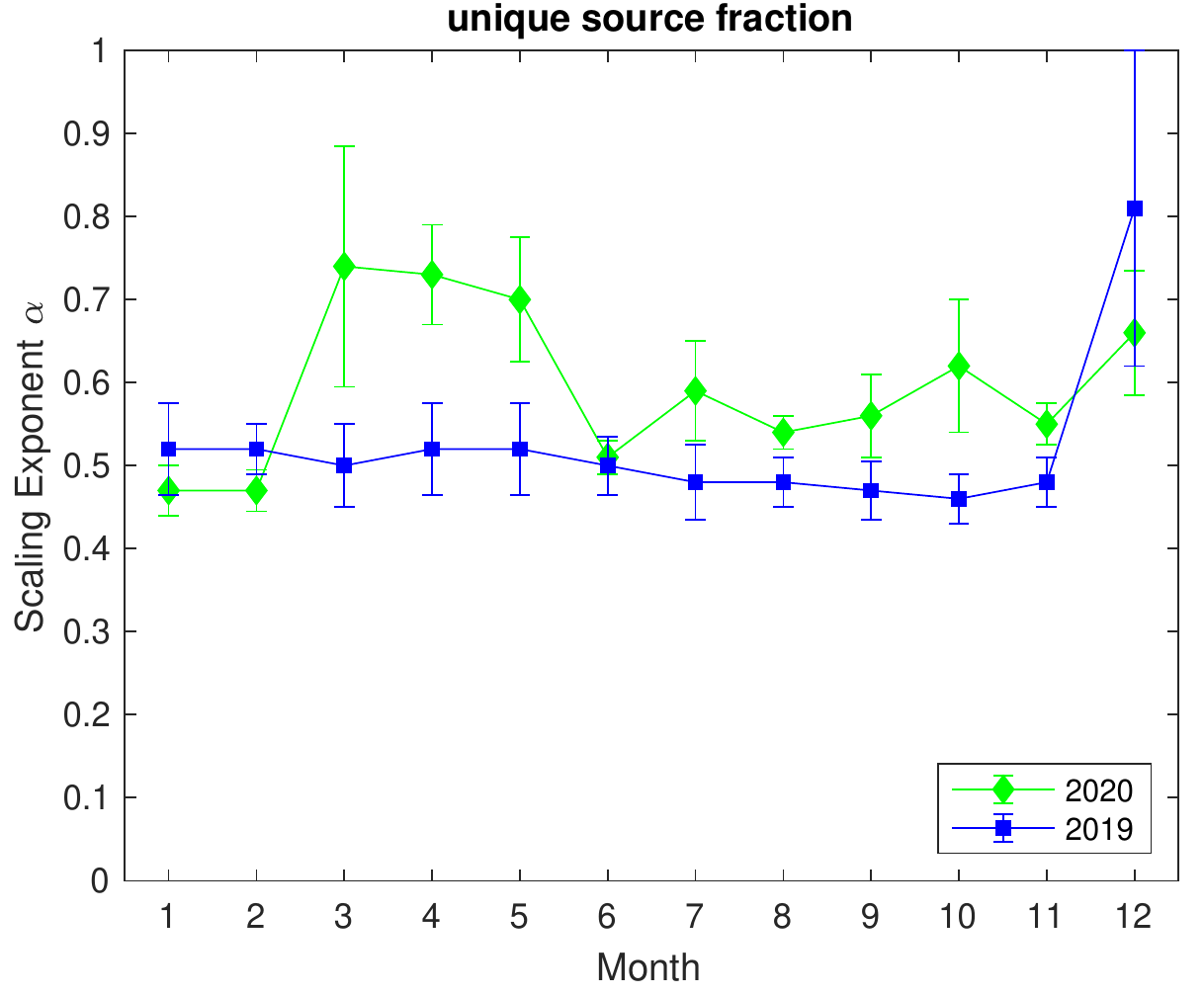}}
\center{\includegraphics[width=0.8\columnwidth]{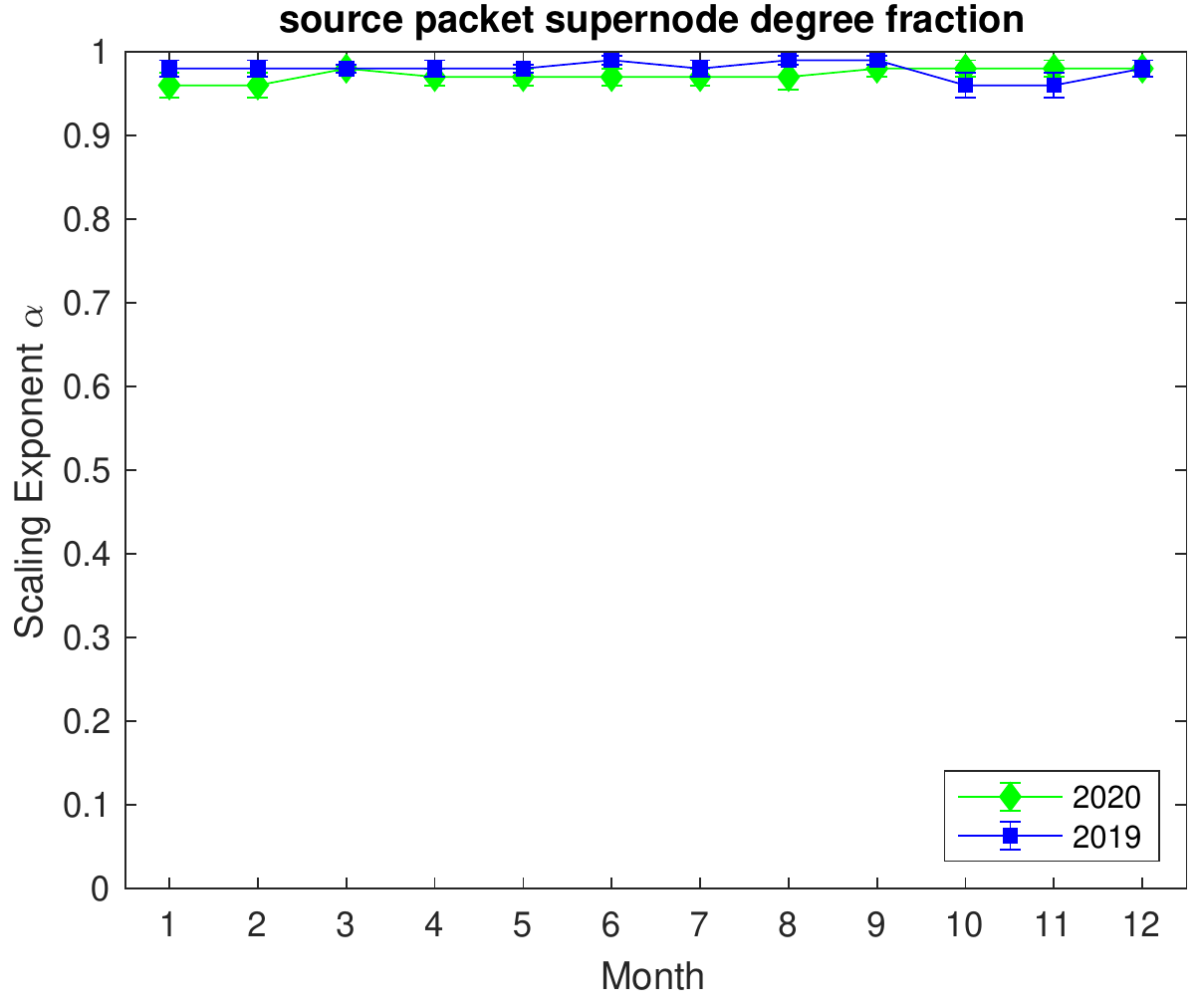}}
\center{\includegraphics[width=0.8\columnwidth]{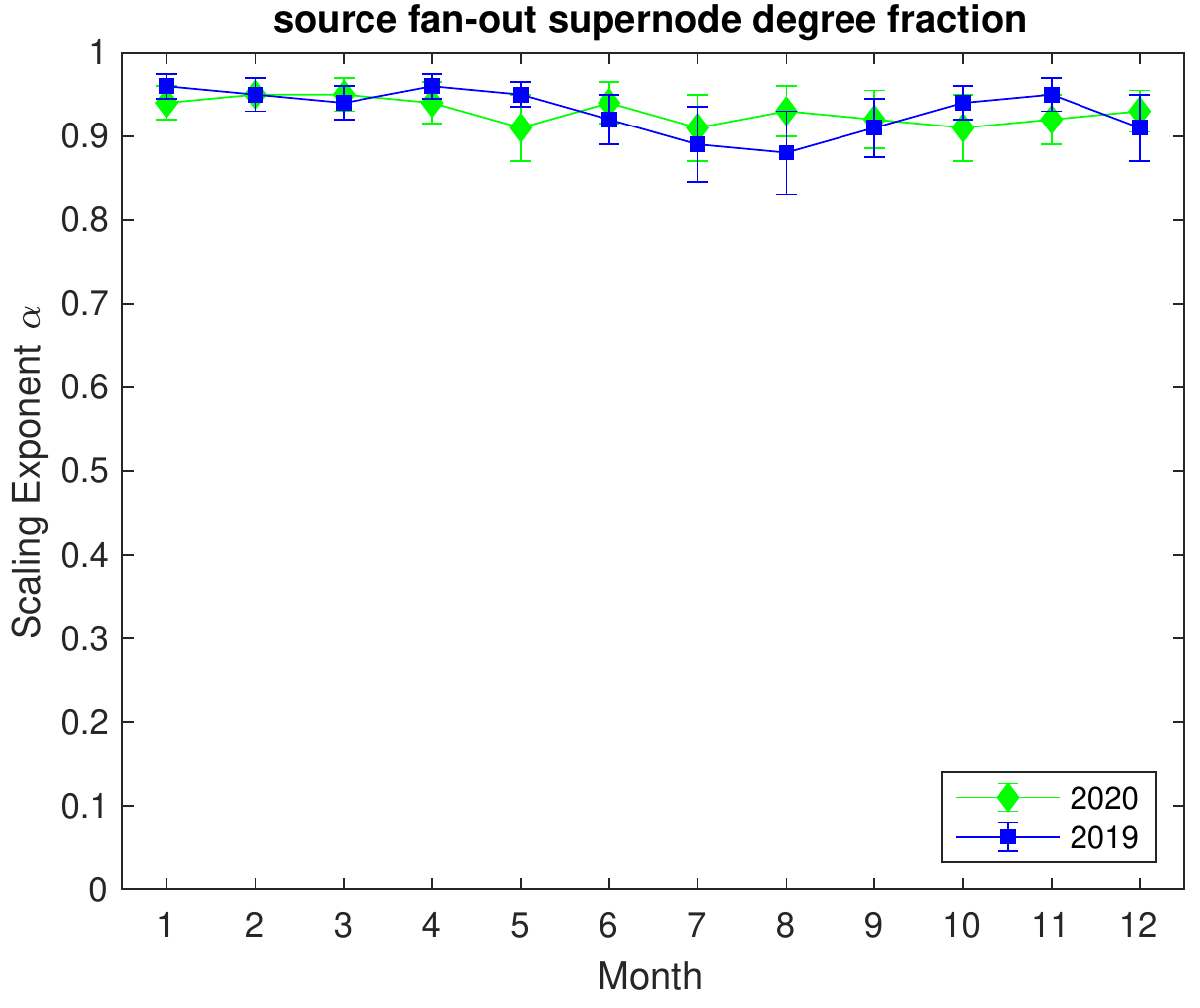}}
      	\caption{{\bf Source scaling exponents}.}
      	\label{fig:SourceScaling}
\end{figure}

\begin{figure}
\center{\includegraphics[width=0.8\columnwidth]{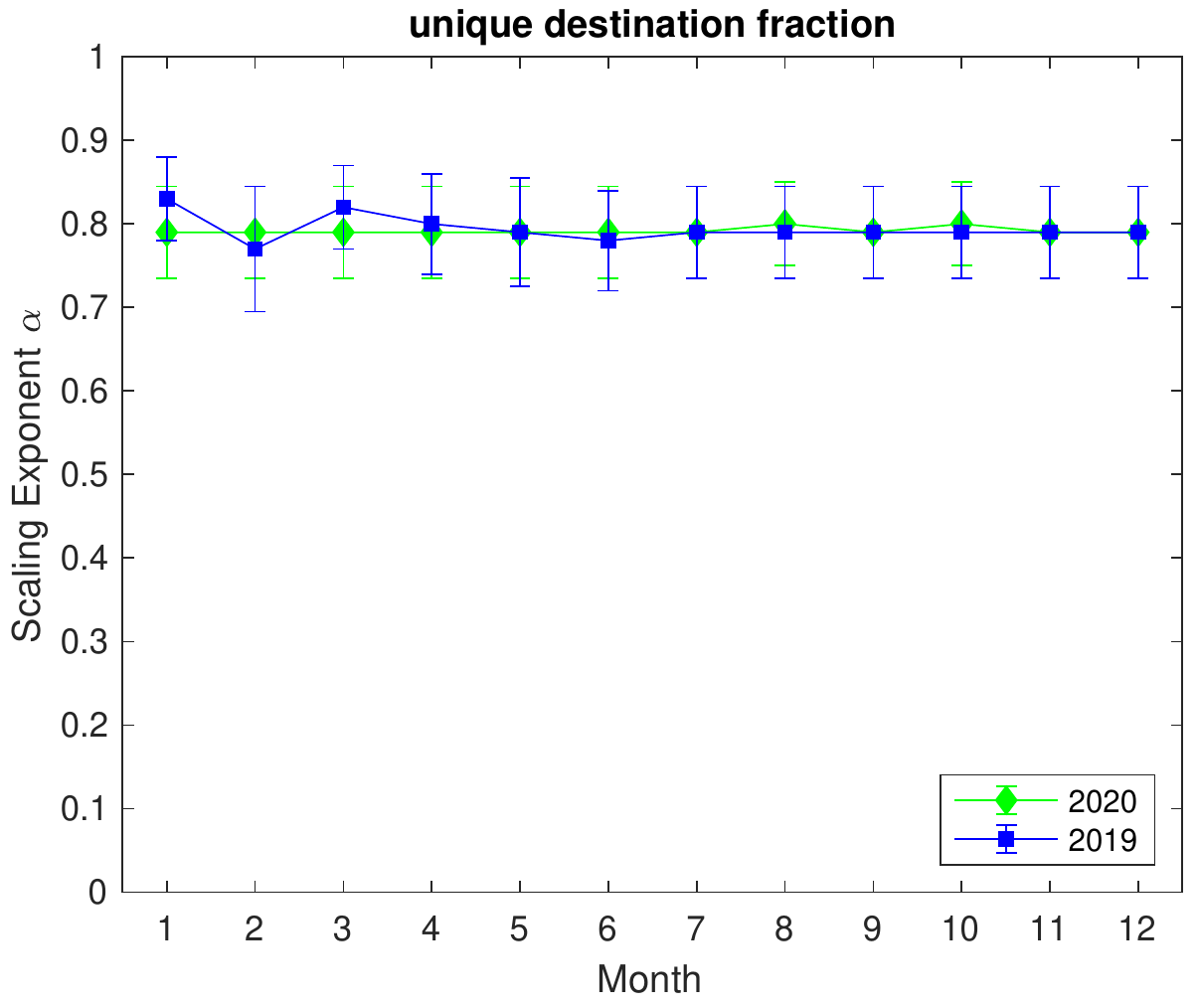}}
\center{\includegraphics[width=0.8\columnwidth]{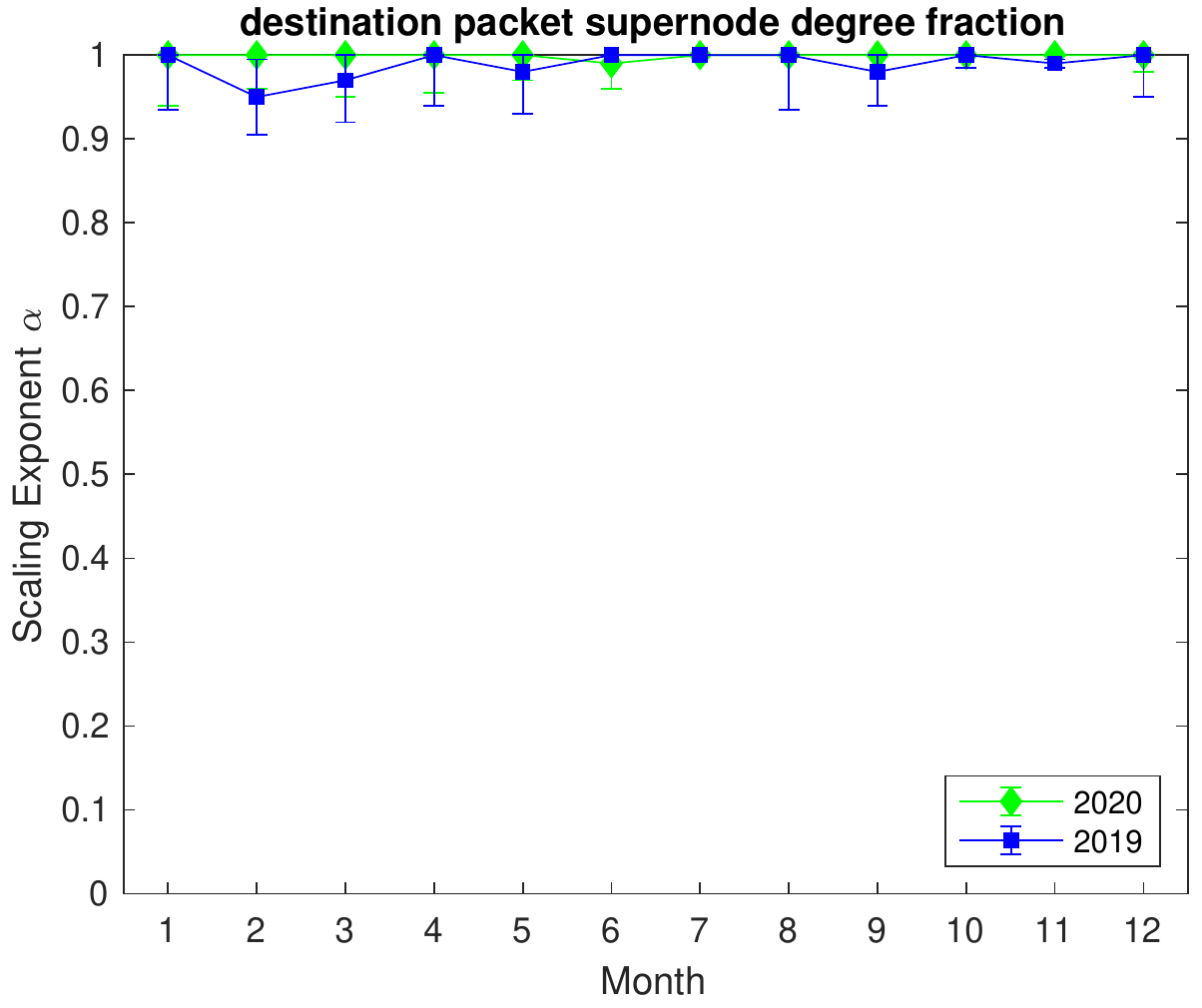}}
\center{\includegraphics[width=0.8\columnwidth]{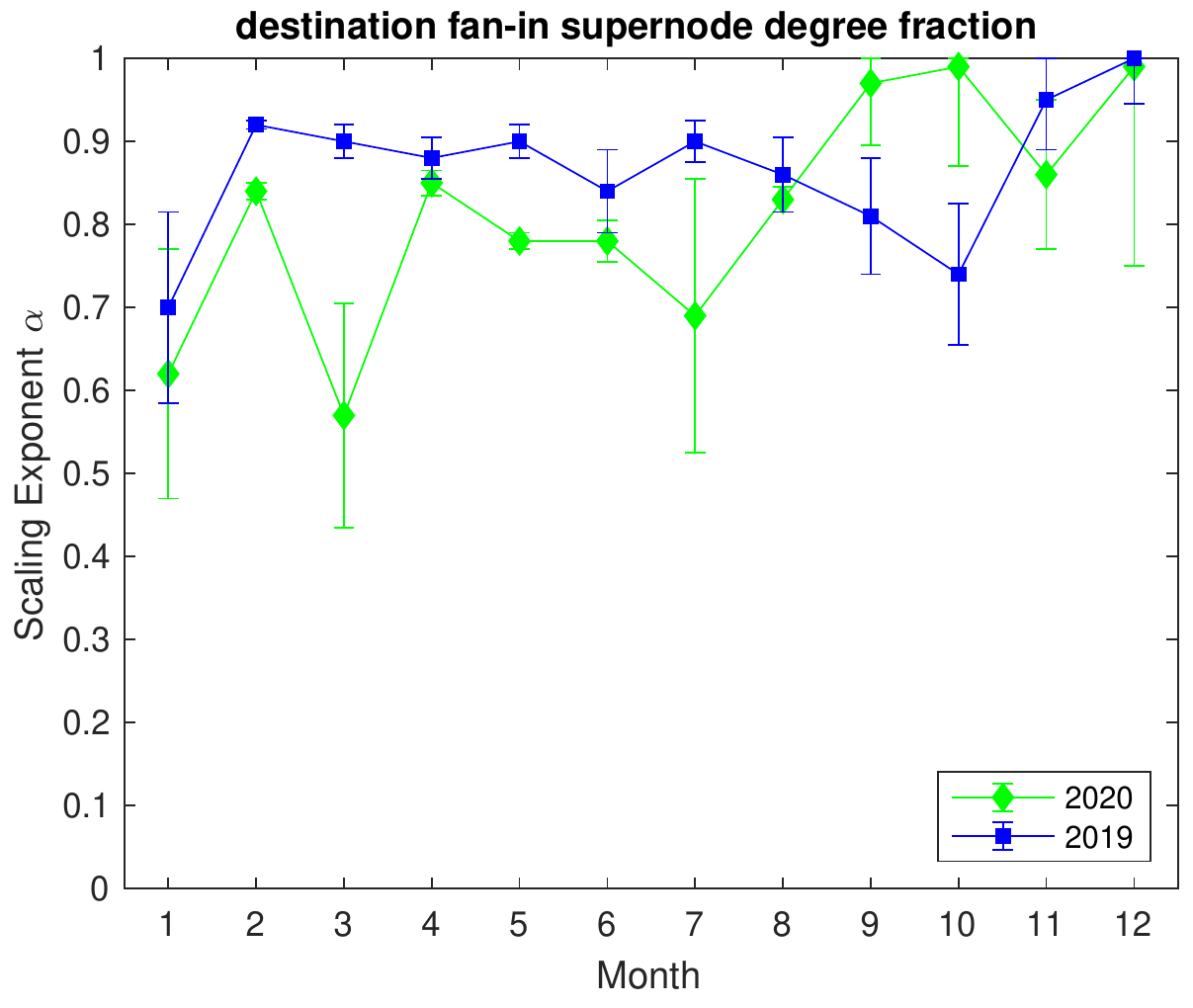}}
      	\caption{{\bf Destination scaling exponents}.}
      	\label{fig:DestinationScaling}
\end{figure}

\begin{figure}
\center{\includegraphics[width=0.8\columnwidth]{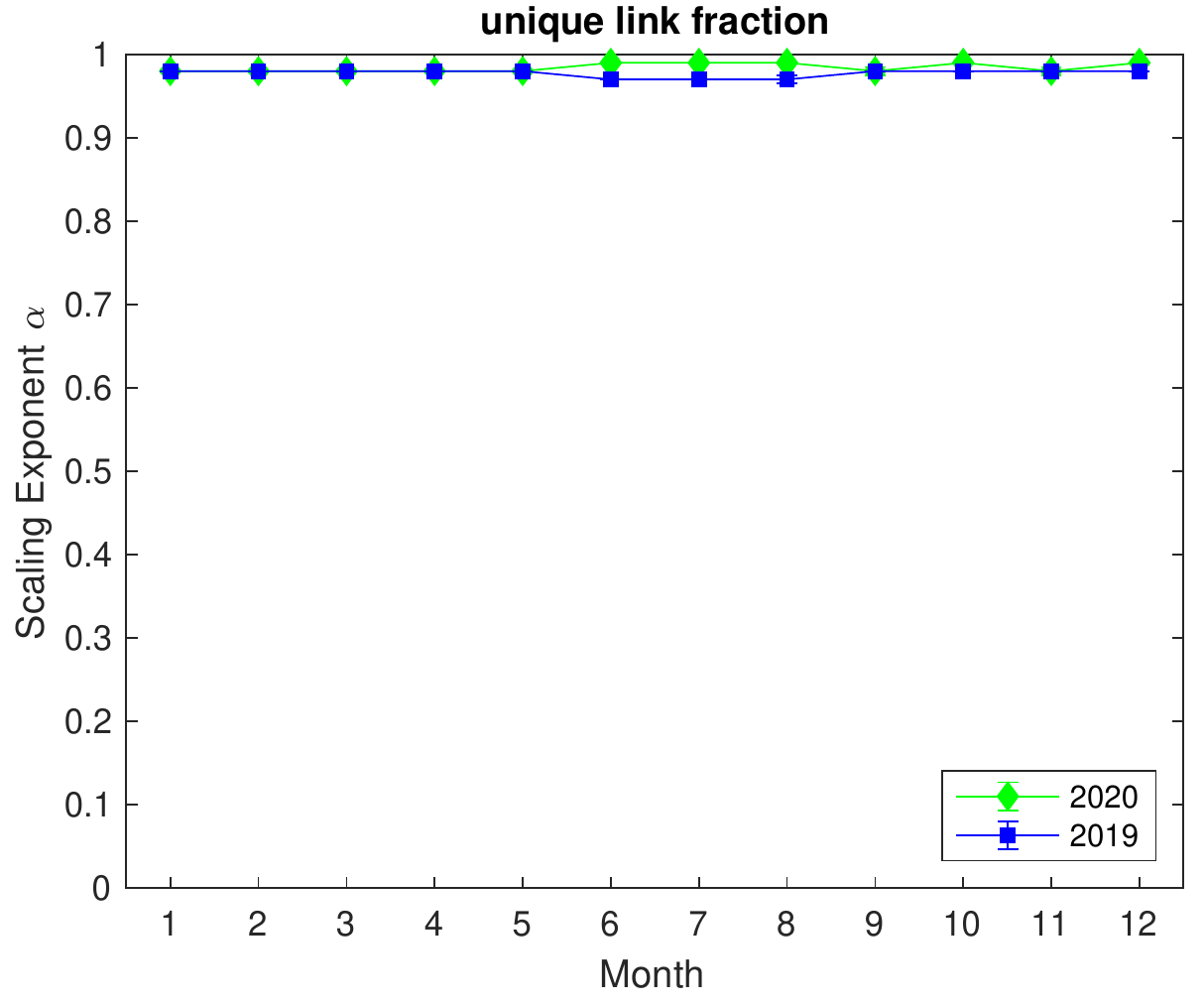}}
\center{\includegraphics[width=0.8\columnwidth]{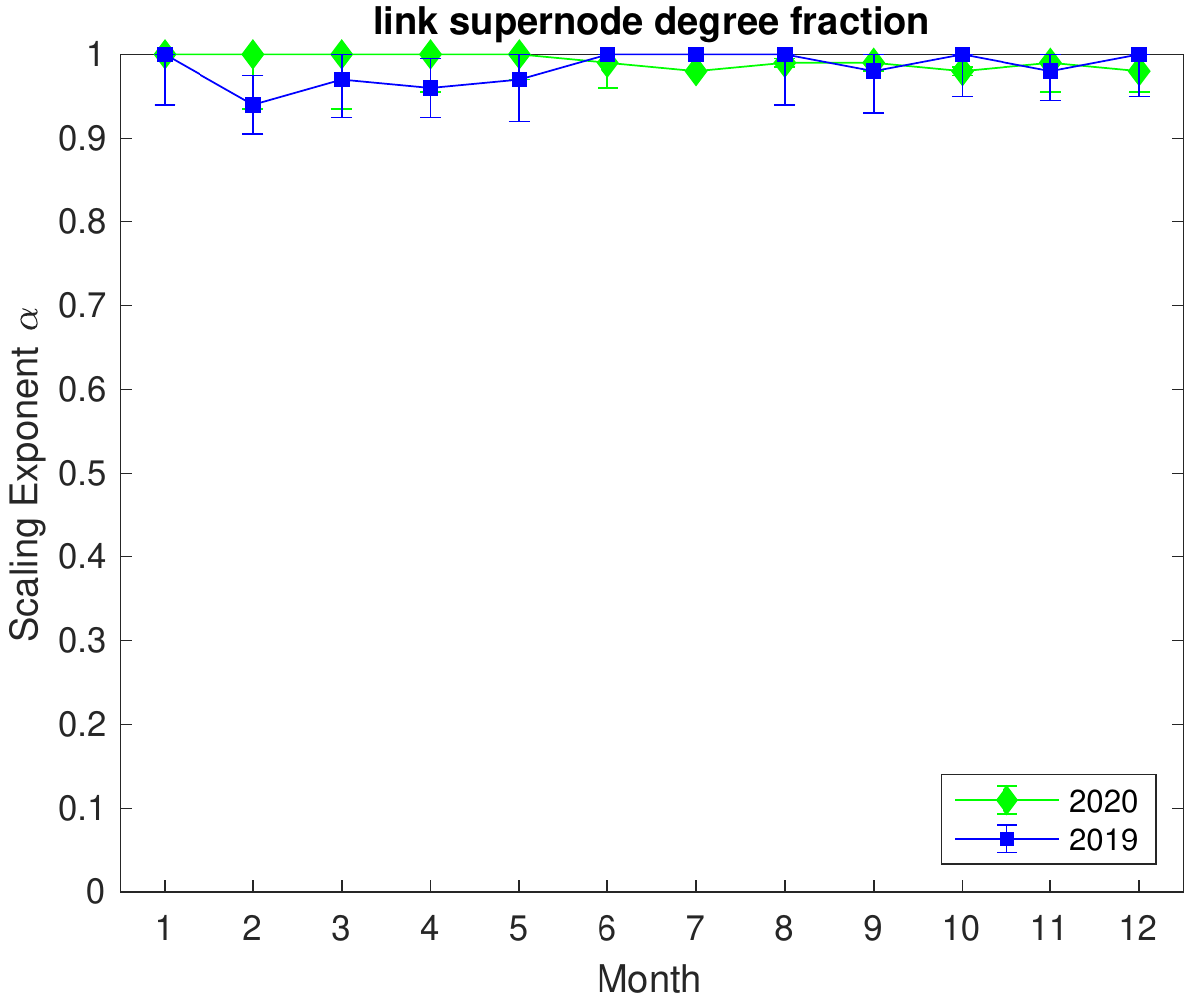}}
      	\caption{{\bf Link scaling exponents}.}
      	\label{fig:LinkScaling}
\end{figure}

Over 40,000,000,000,000 CAIDA Telescope anonymized packet headers from 2019 and 2020 have been collected and analyzed.  The network quantities in Table~\ref{tab:Aggregates} have been computed for window sizes corresponding to
$$
N_V = 2^{17}, 2^{18}, 2^{19}, 2^{20}, 2^{21}, 2^{22}, 2^{23}, 2^{24}, 2^{25}, 2^{26}, 2^{27}
$$
The averages and standard deviations of these quantities have been computed over sets of $2^{27}$ packets.  A key property is how the various network quantities scale as a function of the window size $N_V$.  Figure~\ref{fig:UniqueSourceFraction} (top) shows the average total number of unique sources as a fraction of the packet window $N_V$ measured for the months of 2019-06 and 2020-06.  Figure~\ref{fig:UniqueSourceFraction} (bottom) is the result of scaling the data using the formula
$$
\beta N_{V}^\alpha
$$

Performing a similar analysis across all the network quantities produced the scaling relations for each month.  Figures~\ref{fig:SourceScaling}, \ref{fig:DestinationScaling}, and \ref{fig:LinkScaling}  show the scaling exponents $\alpha$ for the sources, destinations, and links.  In most cases, these scaling exponents are remarkably consistent and lie in the range $0.9 < \alpha < 1.0$.  Three notable exceptions are the scaling of the unique sources, the unique destinations, and the destination supernode fan-in (destination with the most unique sources).  The unique sources shown in Figure~\ref{fig:SourceScaling} (top) appear to scale as $N_V^{0.5}$ in 2019, which increases to $N_V^{0.7}$ appearing in 2020, implying that the relative diversity of observed darkspace sources grew during 2020.  The unique destinations shown in Figure~\ref{fig:DestinationScaling} (top) consistently scaled as $N_V^{0.8}$ throughout 2019 and 2020 indicating that although the source variety may have increased the set of destination addresses they were reaching out to remained similar.  The destination supernode fan-in scaling shows significant fluctuations throughout 2019 and 2020 spanning $0.5 < \alpha < 1.0$.

Table~\ref{tab:ScalingRelations} summarizes the median scaling relations of both the averages and standard deviations of all the network quantities for 2019 and 2020.  These results reveal a strong dependence on these quantities as a function of the packet window size $N_V$ as well as remarkable consistency between 2019 and 2020. To our knowledge, these scaling relations have not been previously reported and represent a new view on the background behavior of network traffic.  The scaling relations provide a new baseline for predicting and reasoning about the nature of this traffic.

\section{Conclusions and Future Work}

Understanding the behavior of the Internet is essential as the Internet has never been more important to our society.  The CAIDA Telescope provides a unique perspective on the Internet by observing a continuous stream of darkspace traffic representing 1/256 of the Internet.  Over 40,000,000,000,000 unique packets were collected during 2019 and 2020.  This is the largest public corpus of Internet traffic ever collected. The Supercomputing Centers at UC San Diego, Lawrence Berkeley National Laboratory, and MIT have combined their resources to analyze the spatial and temporal structure of anonymized source-destination pairs leveraging GraphBLAS hierarchical hypersparse matrices.  Analysis of this unsolicited Internet darkspace traffic has revealed many previously unseen scaling relations.  The data show a significant sustained increase in unsolicited traffic corresponding to the start of the COVID19 pandemic, but relatively little change in the underlying scaling relations associated with unique sources, source fan-outs, unique links, destination fan-ins, and unique destinations.  This work provides a demonstration of the practical feasibility and benefit of the safe collection and analysis of significant of quantities of anonymized Internet traffic.

\section*{Acknowledgments}

The authors wish to acknowledge the following individuals for their contributions and support: Bob Bond, Ronisha Carter, Cary Conrad, Alan Edelman, Tucker Hamilton, Jeff Gottschalk, Nathan Frey, Chris Hill, Hayden Jananthan, Mike Kanaan, Tim Kraska, Andrew Morris, Charles Leiserson, Dave Martinez, Mimi McClure, Joseph McDonald, Christian Prothmann, John Radovan, Steve Rejto, Daniela Rus, Allan Vanterpool, Matthew Weiss, Marc Zissman.

\bibliographystyle{ieeetr}
\bibliography{StreamingCADIAanalysis}

\begin{thebibliography}{10}

\bibitem{Cisco2018-2023}
``{\it Cisco Visual Networking Index: Forecast and Trends, 2018–2023}.''
  https://www.cisco.com/c/en/us/solutions/collateral/executive-perspectives/annual-internet-report/white-paper-c11-741490.html.

\bibitem{kepner2021zero}
J.~Kepner, J.~Bernays, S.~Buckley, K.~Cho, C.~Conrad, L.~Daigle, K.~Erhardt,
  V.~Gadepally, B.~Greene, M.~Jones, R.~Knake, B.~Maggs, P.~Michaleas,
  C.~Meiners, A.~Morris, A.~Pentland, S.~Pisharody, S.~Powazek, A.~Prout,
  P.~Reiner, K.~Suzuki, K.~Takhashi, T.~Tauber, L.~Walker, and D.~Stetson,
  ``Zero botnets: An observe-pursue-counter approach.'' Belfer Center Reports,
  6 2021.

\bibitem{claffy2000measuring}
K.~Claffy, ``Measuring the internet,'' {\em IEEE Internet Computing}, vol.~4,
  no.~1, pp.~73--75, 2000.

\bibitem{li2013survey}
B.~Li, J.~Springer, G.~Bebis, and M.~H. Gunes, ``A survey of network flow
  applications,'' {\em Journal of Network and Computer Applications}, vol.~36,
  no.~2, pp.~567--581, 2013.

\bibitem{rabinovich2016measuring}
M.~Rabinovich and M.~Allman, ``Measuring the internet,'' {\em IEEE Internet
  Computing}, vol.~20, no.~4, pp.~6--8, 2016.

\bibitem{ClaffyClark2020}
k.~claffy and D.~Clark, ``Workshop on internet economics (wie 2019) report,''
  {\em SIGCOMM Comput. Commun. Rev.}, vol.~50, p.~53–59, May 2020.

\bibitem{lumsdaine2007challenges}
A.~Lumsdaine, D.~Gregor, B.~Hendrickson, and J.~Berry, ``Challenges in parallel
  graph processing,'' {\em Parallel Processing Letters}, vol.~17, no.~01,
  pp.~5--20, 2007.

\bibitem{kolda2009tensor}
T.~G. Kolda and B.~W. Bader, ``Tensor decompositions and applications,'' {\em
  SIAM review}, vol.~51, no.~3, pp.~455--500, 2009.

\bibitem{hilbert2011world}
M.~Hilbert and P.~L{\'o}pez, ``The world's technological capacity to store,
  communicate, and compute information,'' {\em Science}, p.~1200970, 2011.

\bibitem{Kepner2009}
J.~Kepner, {\em Parallel MATLAB for Multicore and Multinode Computers}.
\newblock SIAM, 2009.

\bibitem{kepner2011graph}
J.~Kepner and J.~Gilbert, {\em Graph algorithms in the language of linear
  algebra}.
\newblock SIAM, 2011.

\bibitem{kepner2018mathematics}
J.~Kepner and H.~Jananthan, {\em Mathematics of big data: Spreadsheets,
  databases, matrices, and graphs}.
\newblock MIT Press, 2018.

\bibitem{reuther2018interactive}
A.~{Reuther}, J.~{Kepner}, C.~{Byun}, S.~{Samsi}, W.~{Arcand}, D.~{Bestor},
  B.~{Bergeron}, V.~{Gadepally}, M.~{Houle}, M.~{Hubbell}, M.~{Jones},
  A.~{Klein}, L.~{Milechin}, J.~{Mullen}, A.~{Prout}, A.~{Rosa}, C.~{Yee}, and
  P.~{Michaleas}, ``Interactive supercomputing on 40,000 cores for machine
  learning and data analysis,'' in {\em 2018 IEEE High Performance extreme
  Computing Conference (HPEC)}, pp.~1--6, 2018.

\bibitem{gadepally2018hyperscaling}
V.~{Gadepally}, J.~{Kepner}, L.~{Milechin}, W.~{Arcand}, D.~{Bestor},
  B.~{Bergeron}, C.~{Byun}, M.~{Hubbell}, M.~{Houle}, M.~{Jones},
  P.~{Michaleas}, J.~{Mullen}, A.~{Prout}, A.~{Rosa}, C.~{Yee}, S.~{Samsi}, and
  A.~{Reuther}, ``Hyperscaling internet graph analysis with d4m on the mit
  supercloud,'' in {\em 2018 IEEE High Performance extreme Computing Conference
  (HPEC)}, pp.~1--6, Sep. 2018.

\bibitem{kepner19streaming}
J.~{Kepner}, V.~{Gadepally}, L.~{Milechin}, S.~{Samsi}, W.~{Arcand},
  D.~{Bestor}, W.~{Bergeron}, C.~{Byun}, M.~{Hubbell}, M.~{Houle}, M.~{Jones},
  A.~{Klein}, P.~{Michaleas}, J.~{Mullen}, A.~{Prout}, A.~{Rosa}, C.~{Yee}, and
  A.~{Reuther}, ``Streaming 1.9 billion hypersparse network updates per second
  with d4m,'' in {\em 2019 IEEE High Performance Extreme Computing Conference
  (HPEC)}, pp.~1--6, 2019.

\bibitem{kepner202075}
J.~Kepner, T.~Davis, C.~Byun, W.~Arcand, D.~Bestor, W.~Bergeron, V.~Gadepally,
  M.~Hubbell, M.~Houle, M.~Jones, A.~Klein, P.~Michaleas, L.~Milechin,
  J.~Mullen, A.~Prout, A.~Rosa, S.~Samsi, C.~Yee, and A.~Reuther,
  ``75,000,000,000 streaming inserts/second using hierarchical hypersparse
  graphblas matrices,'' in {\em 2020 IEEE International Parallel and
  Distributed Processing Symposium Workshops (IPDPSW)}, pp.~207--210, 2020.

\bibitem{kepner19hypersparse}
J.~{Kepner}, K.~{Cho}, K.~{Claffy}, V.~{Gadepally}, P.~{Michaleas}, and
  L.~{Milechin}, ``Hypersparse neural network analysis of large-scale internet
  traffic,'' in {\em 2019 IEEE High Performance Extreme Computing Conference
  (HPEC)}, pp.~1--11, 2019.

\bibitem{kepner2020multi}
J.~Kepner, C.~Meiners, C.~Byun, S.~McGuire, T.~Davis, W.~Arcand, J.~Bernays,
  D.~Bestor, W.~Bergeron, V.~Gadepally, R.~Harnasch, M.~Hubbell, M.~Houle,
  M.~Jones, A.~Kirby, A.~Klein, L.~Milechin, J.~Mullen, A.~Prout, A.~Reuther,
  A.~Rosa, S.~Samsi, D.~Stetson, A.~Tse, C.~Yee, and P.~Michaleas,
  ``Multi-temporal analysis and scaling relations of 100,000,000,000 network
  packets,'' in {\em 2020 IEEE High Performance Extreme Computing Conference
  (HPEC)}, pp.~1--6, 2020.

\bibitem{devlin2021hybrid}
P.~Devlin, J.~Kepner, A.~Luo, and E.~Meger, ``Hybrid power-law models of
  network traffic,'' {\em arXiv preprint arXiv:2103.15928}, 2021.

\bibitem{huang2018software}
D.~Huang, A.~Chowdhary, and S.~Pisharody, {\em Software-Defined networking and
  security: from theory to practice}.
\newblock CRC Press, 2018.

\bibitem{kepner16mathematical}
J.~{Kepner}, P.~{Aaltonen}, D.~{Bader}, A.~{Bulu{\c{c}}}, F.~{Franchetti},
  J.~{Gilbert}, D.~{Hutchison}, M.~{Kumar}, A.~{Lumsdaine}, H.~{Meyerhenke},
  S.~{McMillan}, C.~{Yang}, J.~D. {Owens}, M.~{Zalewski}, T.~{Mattson}, and
  J.~{Moreira}, ``Mathematical foundations of the graphblas,'' in {\em 2016
  IEEE High Performance Extreme Computing Conference (HPEC)}, pp.~1--9, 2016.

\bibitem{buluc17design}
A.~{Bulu{\c{c}}}, T.~{Mattson}, S.~{McMillan}, J.~{Moreira}, and C.~{Yang},
  ``Design of the graphblas api for c,'' in {\em 2017 IEEE International
  Parallel and Distributed Processing Symposium Workshops (IPDPSW)},
  pp.~643--652, 2017.

\bibitem{davis18algorithm}
T.~A. Davis, ``Algorithm 1000: Suitesparse:graphblas: Graph algorithms in the
  language of sparse linear algebra,'' {\em ACM Trans. Math. Softw.}, vol.~45,
  Dec. 2019.

\bibitem{fan2004prefix}
J.~Fan, J.~Xu, M.~H. Ammar, and S.~B. Moon, ``Prefix-preserving ip address
  anonymization: measurement-based security evaluation and a new
  cryptography-based scheme,'' {\em Computer Networks}, vol.~46, no.~2,
  pp.~253--272, 2004.

\bibitem{byun2019optimizing}
C.~Byun, J.~Kepner, W.~Arcand, D.~Bestor, W.~Bergeron, M.~Hubbell,
  V.~Gadepally, M.~Houle, M.~Jones, A.~Klein, L.~Milechin, P.~Michaleas,
  J.~Mullen, A.~Prout, A.~Rosa, S.~Samsi, C.~Yee, and A.~Reuther, ``Optimizing
  xeon phi for interactive data analysis,'' in {\em 2019 IEEE High Performance
  Extreme Computing Conference (HPEC)}, pp.~1--6, 2019.

\bibitem{byun2019large}
C.~Byun, J.~Kepner, W.~Arcand, D.~Bestor, B.~Bergeron, V.~Gadepally, M.~Houle,
  M.~Hubbell, M.~Jones, A.~Klein, P.~Michaleas, J.~Mullen, A.~Prout, A.~Rosa,
  S.~Samsi, C.~Yee, and A.~Reuther, ``Large scale parallelization using
  file-based communications,'' in {\em 2019 IEEE High Performance Extreme
  Computing Conference (HPEC)}, pp.~1--7, 2019.

\bibitem{elmore2015demonstration}
A.~J. Elmore, J.~Duggan, M.~Stonebraker, M.~Balazinska, U.~Cetintemel,
  V.~Gadepally, J.~Heer, B.~Howe, J.~Kepner, T.~Kraska, {\em et~al.}, ``A
  demonstration of the bigdawg polystore system,'' {\em Proceedings of the VLDB
  Endowment}, vol.~8, no.~12, p.~1908, 2015.

\bibitem{kraska18case}
T.~Kraska, A.~Beutel, E.~H. Chi, J.~Dean, and N.~Polyzotis, ``The case for
  learned index structures,'' in {\em Proceedings of the 2018 International
  Conference on Management of Data}, SIGMOD 18, (New York, NY, USA),
  pp.~489--504, Association for Computing Machinery, 2018.

\bibitem{do20classifying}
E.~H. {Do} and V.~N. {Gadepally}, ``Classifying anomalies for network
  security,'' in {\em ICASSP 2020 - 2020 IEEE International Conference on
  Acoustics, Speech and Signal Processing (ICASSP)}, pp.~2907--2911, 2020.

\bibitem{soule2004identify}
A.~Soule, A.~Nucci, R.~Cruz, E.~Leonardi, and N.~Taft, ``How to identify and
  estimate the largest traffic matrix elements in a dynamic environment,'' in
  {\em ACM SIGMETRICS Performance Evaluation Review}, vol.~32, pp.~73--84, ACM,
  2004.

\bibitem{zhang2005estimating}
Y.~Zhang, M.~Roughan, C.~Lund, and D.~L. Donoho, ``Estimating point-to-point
  and point-to-multipoint traffic matrices: an information-theoretic
  approach,'' {\em IEEE/ACM Transactions on Networking (TON)}, vol.~13, no.~5,
  pp.~947--960, 2005.

\bibitem{mucha2010community}
P.~J. Mucha, T.~Richardson, K.~Macon, M.~A. Porter, and J.-P. Onnela,
  ``Community structure in time-dependent, multiscale, and multiplex
  networks,'' {\em science}, vol.~328, no.~5980, pp.~876--878, 2010.

\bibitem{tune2013internet}
P.~Tune, M.~Roughan, H.~Haddadi, and O.~Bonaventure, ``Internet traffic
  matrices: A primer,'' {\em Recent Advances in Networking}, vol.~1, pp.~1--56,
  2013.

\bibitem{karvanen2003measuring}
J.~Karvanen and A.~Cichocki, ``Measuring sparseness of noisy signals,'' in {\em
  4th International Symposium on Independent Component Analysis and Blind
  Signal Separation}, pp.~125--130, 2003.

\bibitem{clauset2009power}
A.~Clauset, C.~R. Shalizi, and M.~E. Newman, ``Power-law distributions in
  empirical data,'' {\em SIAM review}, vol.~51, no.~4, pp.~661--703, 2009.

\bibitem{barabasi2016network}
A.-L. Barab{\'a}si {\em et~al.}, {\em Network science}.
\newblock Cambridge university press, 2016.

\bibitem{leland1994self}
W.~E. Leland, M.~S. Taqqu, W.~Willinger, and D.~V. Wilson, ``On the
  self-similar nature of ethernet traffic (extended version),'' {\em IEEE/ACM
  Transactions on Networking (ToN)}, vol.~2, no.~1, pp.~1--15, 1994.

\bibitem{faloutsos1999power}
M.~Faloutsos, P.~Faloutsos, and C.~Faloutsos, ``On power-law relationships of
  the internet topology,'' in {\em ACM SIGCOMM computer communication review},
  vol.~29-4, pp.~251--262, ACM, 1999.

\bibitem{albert1999internet}
R.~Albert, H.~Jeong, and A.-L. Barab{\'a}si, ``Internet: Diameter of the
  world-wide web,'' {\em Nature}, vol.~401, no.~6749, p.~130, 1999.

\bibitem{barabasi1999emergence}
A.-L. Barab{\'a}si and R.~Albert, ``Emergence of scaling in random networks,''
  {\em Science}, vol.~286, no.~5439, pp.~509--512, 1999.

\bibitem{adamic2000power}
L.~A. Adamic and B.~A. Huberman, ``Power-law distribution of the world wide
  web,'' {\em science}, vol.~287, no.~5461, pp.~2115--2115, 2000.

\bibitem{barabasi2009scale}
A.-L. Barab{\'a}si, ``Scale-free networks: a decade and beyond,'' {\em
  science}, vol.~325, no.~5939, pp.~412--413, 2009.

\bibitem{mahanti2013tale}
A.~Mahanti, N.~Carlsson, A.~Mahanti, M.~Arlitt, and C.~Williamson, ``A tale of
  the tails: Power-laws in internet measurements,'' {\em IEEE Network},
  vol.~27, no.~1, pp.~59--64, 2013.

\end{thebibliography}

\appendices
\setcounter{equation}{0}
\renewcommand{\theequation}{\thesection\arabic{equation}}

\end{document}